\documentclass[11pt]{article}
\pdfoutput=1
\usepackage[centertags]{amsmath}
\usepackage[square, comma, sort&compress,numbers]{natbib}
\usepackage{array,multirow}
\numberwithin{equation}{section}
\usepackage{amssymb,amsfonts}
\usepackage{graphicx}
\usepackage{color}
\usepackage{mathtools,bm}
\usepackage{accents}
\usepackage[dvipsnames]{xcolor}

\usepackage{epsfig}
\usepackage{bbold}

\usepackage{wrapfig}
\usepackage{float}
\usepackage{soul}
\usepackage{tkz-euclide}
\usepackage{braket}

\usepackage{tikz,pgf}
\usetikzlibrary{shapes}
\usetikzlibrary{calc}
\usetikzlibrary{decorations.pathmorphing}
\usetikzlibrary{decorations.pathreplacing,shapes.misc}
\usetikzlibrary{positioning}
\usetikzlibrary{arrows}
\usetikzlibrary{decorations.markings}
\usetikzlibrary{shadings}

\usetikzlibrary{intersections}

\def\be{\begin{equation}}
\def\ee{\end{equation}}
\def\ba{\begin{array}}
\def\ea{\end{array}}

\def\bst{\begin{split}}
\def\est{\end{split}}

\def\dps{\displaystyle}

\def\Li2{\operatorname{Li_2}}

\newdimen\tableauside\tableauside=1.0ex
\newdimen\tableaurule\tableaurule=0.4pt
\newdimen\tableaustep
\def\phantomhrule#1{\hbox{\vbox to0pt{\hrule height\tableaurule
width#1\vss}}}
\def\phantomvrule#1{\vbox{\hbox to0pt{\vrule width\tableaurule
height#1\hss}}}
\def\sqr{\vbox{%
\phantomhrule\tableaustep

\hbox{\phantomvrule\tableaustep\kern\tableaustep\phantomvrule\tableaustep}%
\hbox{\vbox{\phantomhrule\tableauside}\kern-\tableaurule}}}
\def\squares#1{\hbox{\count0=#1\noindent\loop\sqr
\advance\count0 by-1 \ifnum\count0>0\repeat}}
\def\tableau#1{\vcenter{\offinterlineskip
\tableaustep=\tableauside\advance\tableaustep by-\tableaurule
\kern\normallineskip\hbox
{\kern\normallineskip\vbox
{\gettableau#1 0 }%
\kern\normallineskip\kern\tableaurule}%
\kern\normallineskip\kern\tableaurule}}
\def\gettableau#1 {\ifnum#1=0\let\next=\null\else
\squares{#1}\let\next=\gettableau\fi\next}

\newcommand{\bref}[1]{\textbf{\ref{#1}}}

\newcommand{\RR}{\mathbb{R}}

\newcommand{\ZZ}{\mathbb{Z}}

\def\cA{\mathcal{A}}
\def\cB{\mathcal{B}}
\def\cC{\mathcal{C}}

\def\cG{\mathcal{G}}
\def\cH{\mathcal{H}}
\def\cI{\mathcal{I}}

\def\cS{\mathcal{S}}

\def\cU{\mathcal{U}}
\def\cV{\mathcal{V}}
\def\cW{\mathcal{W}}

\def\rS{{\rm S}}

\def\rV{{\rm V}}

\numberwithin{equation}{section} \makeatletter
\@addtoreset{equation}{section}

\def\be{\begin{equation}}
\def\ee{\end{equation}}
\def\ba{\begin{array}}
\def\ea{\end{array}}

\def\dps{\displaystyle}

\def\ba{\begin{array}}
\def\ea{\end{array}}

\def\dps{\displaystyle}

\def\cft{CFT$_D$ }

\def\hd{\frac{D}{2}}

\newcommand{\intbrom}{\int_{-i \infty}^{+ i \infty}}

\def\C2{\text{C}_2}

\def\id{\mathbb{1}}

\def\BF{\Phi}

\newcommand{\triple}[1]{\langle #1 \rangle} 
\newcommand{\cycle}{\mathrm{C}}

\usepackage{jheppub}
\makeatletter
\def\@fpheader{\vspace{-.1cm}}
\makeatother

\title{\centering{Multipoint conformal integrals in $D$ dimensions. Part I \\ {\Large Bipartite Mellin-Barnes representation and reconstruction} }}

\author{Konstantin\ Alkalaev and}
\author{Semyon\ Mandrygin}

\affiliation{I.E. Tamm Department of Theoretical Physics, \\P.N. Lebedev Physical
Institute, 119991 Moscow, Russia}

\emailAdd{alkalaev@lpi.ru}
\emailAdd{semyon.mandrygin@gmail.com}

\abstract{We propose a systematic approach to calculating $n$-point one-loop parametric conformal integrals in $D$ dimensions which we call the reconstruction procedure. It relies on decomposing a conformal integral over basis functions which are generated from a set of master functions by acting with  the cyclic group $\mathbb{Z}_n$.  In order to identify the master functions we introduce a bipartite Mellin-Barnes representation by means of splitting a given conformal integral into two additive parts, one of which can be evaluated explicitly in terms of multivariate generalized  hypergeometric series. 

For the box and pentagon integrals (i.e. $n=4,5$) we show that a computable part of the bipartite representation  contains all master functions. In particular, this allows us to evaluate the parametric pentagon integral as a sum of ten basis functions generated from two master functions by the cyclic group $\mathbb{Z}_5$. The resulting  expression can be  tested in two ways. First, when one of  propagator powers is set to zero, the pentagon integral is reduced to the known box integral, which is also rederived through the reconstruction procedure. Second, going to the non-parametric case, we reproduce the known expression for the pentagon integral given in terms of logarithms derived earlier within the geometric approach to calculating conformal integrals. 

We conclude  by considering the hexagon  integral ($n=6$) for which we show that those basis functions which follow from the computable part of the bipartite representation are not enough and more basis functions are required. In the second part of our project we will describe a method of  constructing a complete set of master/basis functions in the $n$-point case. }

\begin{document}

\maketitle
\flushbottom

\section{Introduction}

Conformal integrals are an interesting specific class of  integrals that arise in various  areas of quantum field theory \cite{Symanzik:1972wj}. E.g. they  appear in the study of  scattering multi-loop amplitudes, where the corresponding Feynman  integrals respect a dual conformal invariance  \cite{Drummond:2006rz,Drummond:2008vq} as well as in the Fishnet CFT models  \cite{Zamolodchikov:1980mb, Gurdogan:2015csr, Caetano:2016ydc, Kazakov:2018qbr, Chicherin:2017cns, Chicherin:2017frs} (see \cite{Loebbert:2022nfu, Corcoran:2023ljn} for reviews). There is a number of methods of calculating  conformal integrals among which the most fruitful are the geometric approach which treats these integrals as volumes of simplices \cite{Davydychev:1997wa,Mason:2010pg,Schnetz:2010pd,Nandan:2013ip, Bourjaily:2019exo, Ren:2023tuj}, Yangian bootstrap \cite{Loebbert:2019vcj,Duhr:2022pch,Duhr:2024hjf,Loebbert:2024qbw}, Gelfand-Kapranov-Zelevinsky (GKZ) differential equations \cite{Pal:2021llg,Pal:2023kgu, Levkovich-Maslyuk:2024zdy} as well as other approaches which combine various techniques \cite{DelDuca:2011wh,Dixon:2011ng,Paulos:2012qa,Duhr:2023bku,Duhr:2023eld,Prabhu:2023oxv,Prabhu:2024khb,He:2025vqt}. Nonetheless, despite this recent remarkable  progress in calculating  conformal integrals, explicit closed-form expressions for general conformal integrals are not yet known.

On the other hand, a more traditional way to come across conformal integrals is to use the shadow formalism in CFT which represents  correlation functions  as particular integrals possessing  conformal invariance by construction \cite{Ferrara:1972uq,Ferrara:1972kab, Dolan:2000uw, Dolan:2000ut,Dolan:2011dv,SimmonsDuffin:2012uy,Rosenhaus:2018zqn}.  In particular, using the shadow formalism allows one to take advantage of various Feynman integral techniques  for CFT calculations.\footnote{Conformal integrals are also instrumental in CFT on non-trivial manifolds  \cite{Fateev:2011qa,Petkou:2021zhg,Karydas:2023ufs,Alkalaev:2023evp,Alkalaev:2024jxh,Gobeil:2018fzy,Belavin:2024nnw,Belavin:2024mzd}.} In this paper, we instead are inspired by CFT methods for dealing with  conformal integrals. Namely, we develop the idea of  asymptotic analysis of  conformal integrals which is borrowed from the conformal block decomposition of  correlation functions in CFT. Recall that the conformal blocks can be defined by doing OPEs between pairs of primary operators inside a correlation function.  This results in choosing a particular channel, i.e. one fixes an order and closeness  of points, which is known as the OPE limit. This eventually defines a domain of convergence of the conformal block in the coordinate space. Now, when the conformal block  is represented in terms of conformal integrals, the latter are calculated in the  OPE limit to have correct asymptotic behaviour. Other asymptotics of the respective conformal integral describe the so-called shadow blocks \cite{SimmonsDuffin:2012uy,Rosenhaus:2018zqn}. The conformal blocks in other coordinate domains, i.e. in other possible channels, can be obtained by means of analytic continuation formulas.     

Relying on this relation between conformal blocks and  conformal integrals we can formulate the first  point of  our approach to calculating $n$-point conformal integrals in $D$ dimensions: one fixes a convenient coordinate domain  where a conformal integral can be represented in terms of a multivariate power series and then analytically continued onto other domains. The second point  is that there is no need to evaluate conformal integral entirely, but only partially. To this end, one decomposes a  conformal integral into two additive parts which are given by particular   Mellin-Barnes integrals. This is a {\it bipartite} representation. The first term  can be evaluated explicitly in some coordinate domain, while the second one is quite complicated and  has not been computed yet. Here, we put forward our main idea that the second term can be reconstructed from the first one by invoking  permutation invariance of the full conformal integral. Note that  such a partition of  conformal integral in no way is invariant against  transformations from the symmetric group $\cS_n$. Thus, the third point of our approach is to introduce a subgroup of the symmetric group which is claimed to generate the full conformal integral  by acting on  a set of {\it master} functions. We call this calculation scheme a reconstruction. 

The most intricate part of this approach is to identify both a generating group and  master   functions. It turns out that a generating group can be chosen as the cyclic group $\mathbb{Z}_n \subset \cS_n$ though this is not the only choice. A generating group is required to act on master functions to generate a complete set of {\it basis} functions which sum up to the full  conformal integral. It follows that the generating group acts on the basis functions by reshuffling them. To single out an independent set of master functions we claim  that they are not related by cyclic permutations.\footnote{In this paper, a cyclic permutation will be understood only as a cycle of maximum  length $n$, which generates the cyclic group $\mathbb{Z}_n \subset \cS_n$.} In other words, a set of basis functions is split into orbits of $\mathbb{Z}_n$, which, in general, have different lengths.

In the cases of the box and pentagon conformal integrals\footnote{We will be operating with integrals in the coordinate representation. At the same time, we use terminology (i.e. box, pentagon, hexagon)  which is inherited from the dual momentum representation, see e.g. \cite{Drummond:2008vq}.}  the corresponding master functions can be naturally identified by explicitly calculating the first term of the bipartite representation  and then analytically  continuing  the resulting expression onto  some  coordinate domain around a specified point. This procedure  yields the first term as a sum of generalized hypergeometric multivariate functions from  which some functions can be selected as master functions. The cyclic group generates a set of basis functions which define the second part of the bipartite representation thereby giving the full conformal integral. 

However, when considering  the hexagon integral we find out that those basis functions which come from the first part of the bipartite representations are not enough to build the full conformal integral. Nonetheless, we put forward a conjecture how to build a complete set of basis functions which is the subject of our forthcoming paper \cite{Alkalaev:2025zhg}. Essentially, there we claim that the basis functions which we are able to identify  explicitly from the bipartite representation have a number  of systematic properties that will eventually allow us to build a desired complete set.

This paper is organized as follows. In section \bref{sec:conf} we introduce the (non)-parametric  $n$-point  conformal integral and  review the standard calculation method  by Symanzik which is based on using the Mellin-Barnes integrals. Then, in section \bref{sec:MB} we propose a bipartite Mellin-Barnes representation which splits the original conformal integral into two additive parts. Here, we explicitly calculate the first part in some coordinate domain and then discuss the second part. In section \bref{sec:perm} we describe the action of symmetric group and consider the analytic continuation to the other coordinate domain. Section \bref{sec:box} considers the parametric box integral which has been studied in detail   in the  literature and examines our reconstruction approach using this example. Summarizing our study of the box integral, in section \bref{sec:sum} we outline a heuristic procedure of obtaining the full asymptotic expansions in the $n$-point case. Then, in section \bref{sec:penta} we apply this reconstruction machinery to the 5-point conformal integral. Here, we find the parametric pentagon integral as a sum of ten basis functions and then perform a series of consistency checks.  In section \bref{sec:hex} we consider the hexagon integral and discover  that the way of choosing master functions used in the box and pentagon cases is insufficient to construct a complete asymptotic expression because there should be more master functions. Finally, in section \bref{sec:recon},  based on the considered  examples, we formulate and discuss  a conjecture regarding  reconstruction of the $n$-point one-loop conformal integral.  In the concluding section \bref{sec:conclusion} we  summarize our results  and elaborate on perspectives for future work.    

A few appendices contain numerous technical aspects of calculations. Appendix  \bref{app:notations} collects  our notation and conventions used throughout the paper. Appendix  \bref{app:functions} describes  various generalized hypergeometric  multivariate functions, their integral representations and analytic continuation formulas. In appendix  \bref{app:MB_more} we suggest one more Mellin-Barnes representation of the conformal integral which can be useful in practice, e.g. for finding the conformal integral using the computer algebra methods.     Appendix  \bref{app:explicit_5}  collects  explicit expressions for the basis functions which define  the  parametric pentagon  conformal integral.  In particular, in appendix  \bref{app:explicit_52} we  examine  our  representation of the non-parametric (i.e. with unit propagator powers) pentagon integral. Appendix  \bref{app:kin} discusses extended kinematic groups which are particular subgroups in the symmetric group.

\section{Multipoint parametric conformal integrals}
\label{sec:conf}

The following integral over the $D$-dimensional Euclidean space $\RR^{D}$ 
\be
\label{indef}
I_n^{{\bm a}}({\bm x}) = \int_{\mathbb{R}^D} \frac{{\rm d}^D x_0}{\pi^{\hd}} \,\prod_{i=1}^n  X_{0i}^{-a_i}\,,
\qquad X_{ij} \equiv X_{i,j} = (x_i - x_j)^2\,,
\quad x_i \in \RR^D\,,
\ee
where ${\bm a} = \{a_1,a_2,..., a_n \}$, ${\bm x} = \{x_1,x_2,..., x_n \}$,  is called the $n$-point one-loop conformal integral \cite{Symanzik:1972wj} provided that the  propagator powers  $a_i\in \RR$  obey the constraint
\be
\label{confcond}
\sum_{i=1}^n a_i = D \,,
\ee
which guarantees  that $I_n^{{\bm a}}({\bm x})$ transforms covariantly under $O(D+1,1)$ conformal transformations (see e.g. \cite{Corcoran:2023ljn} for details). The integrals with arbitrary $a_i$ subjected to \eqref{confcond} are called {\it parametric}, while those with all $a_i = 1$ in $D=n$ dimensions are called {\it non-parametric}.

\begin{figure}[h]
\centering
\includegraphics[scale=0.8]{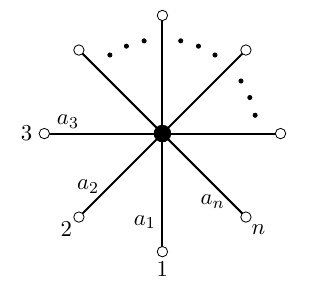}
\caption{The parametric conformal integral  $I_n^{{\bm a}}({\bm x})$ can be depicted as an $n$-valent vertex. The  $i$-th leg denotes the propagator $X_{0i}^{-a_i}$ which is characterized by  $i$-th  position $x_i$ and propagator power $a_i$, the central  dot denotes integration over $x_0$.} 
\label{fig:vertex}
\end{figure}
Since the paper of Symanzik  \cite{Symanzik:1972wj}, the conformal integrals are traditionally represented in terms of the Mellin-Barnes integrals. In this section we  first recall the Symanzik  approach and then introduce a bipartite Mellin-Barnes representation.

\subsection{Review of the Symanzik approach}

In order to calculate the conformal integral \eqref{indef} one can use the Schwinger parametrization 
\be
\label{schwingpar}
\frac{1}{X_{0i}^{a_i}} = \frac{1}{\Gamma(a_i)} \int_{0}^{+\infty} \frac{{\rm d} \lambda_i}{\lambda_i} \lambda_i^{a_i} \exp \left(-\lambda_i X_{0i} \right) \,,
\qquad
a_i > 0\;.
\ee
This standard trick  allows one to evaluate the $D$-dimensional integral over $x_0$ by obtaining a set of one-dimensional integrals over $\lambda_i$. Then, the $n$-point conformal integral is represented as
\be
\label{inlambdas}
I_n^{{\bm a}}({\bm x}) = N_n^{\bm a} \int_0^{{+\infty}} \prod_{i=1}^{n} \left( \frac{{\rm d} \lambda_i}{\lambda_i} \lambda_i^{a_i}  \right) \frac{1}{|\bm \lambda_{1,n}|^{\hd}} \, \operatorname{exp}\Bigg( -\frac{1}{|\bm \lambda_{1,n}|} \sum_{1\leq i < j \leq n} \lambda_i \lambda_j X_{ij} \Bigg)\,,
\ee
where\footnote{The most frequently used notation and conventions are collected in appendix  \bref{app:notations}.} 
 $|\bm \lambda_{i,j}| = \sum_{l=i}^{j} \lambda_l$ and for brevity we defined the prefactor 
\be
\label{N_n}
N_j^{\bm a} = \frac{1}{\Gamma(a_1, ..., a_{j})}\,, \qquad  j = 1,..., n\,,
\ee
for which the standard notation for the product of $\Gamma$-functions \eqref{gammasprod} is used.

The key observation  \cite{Symanzik:1972wj} is that in \eqref{inlambdas} one can substitute $|\bm \lambda_{1,n}| \to \lambda_n$  as a consequence of the conformality constraint  \eqref{confcond}. Representing the  exponential functions by means of the Mellin-Barnes formula \eqref{mbexponent}  
\be
\begin{split}
\operatorname{exp}\left(-\frac{\lambda_i \lambda_j X_{ij}}{\lambda_n} \right) &=  \intbrom \widehat{{\rm d}s}_{ij}  \left(\frac{\lambda_i \lambda_j X_{ij}}{\lambda_n} \right)^{s_{ij}}\,, \qquad\;  1\leq i < j \leq n-2\,, \\ 
\operatorname{exp}\left(-\frac{\lambda_k \lambda_{n-1} X_{k,n-1}}{\lambda_n} \right) &=  \intbrom \widehat{{\rm d}t}_{k}  \left(\frac{\lambda_k \lambda_{n-1} X_{k,n-1}}{\lambda_n} \right)^{t_{k}}\,, \qquad\; 1\leq k \leq n-3 \,, 
\end{split}
\ee
one can successively calculate the integrals over $\lambda_n\,, \lambda_{n-1}\,, ..., \lambda_1$ (the integration measures are defined in \eqref{mb_measure}). As a result, the conformal integral can be represented as a product 
\be
\label{intripleprod}
I_n^{{\bm a}}({\bm x}) = L_n^{{\bm a}}({\bm x}) \, \cI_n^{{\bm a}}({\bm \eta})\,,
\ee
where the {\it leg-factor}  
\be
\label{inleg}
L_n^{{\bm a}}({\bm x}) = \left(Z_{n}^{n-2,n-1} \right)^{-a'_n}  \prod_{i=1}^{n-1}\left( X_{i,n}^{-a_i} \right)  \,, 
\qquad 
a'_i = \hd - a_i\,,
\ee
depends on the following combinations of squared distances,     
\be
\label{zdef}
Z_k^{ij} \equiv Z_k^{i,j} = \frac{X_{ij}}{X_{ik} X_{jk}}\,, 
\qquad 
i,j \neq k\,,
\ee
which conveniently define the cross-ratios as\footnote{\label{fn:cross} The cross-ratios here are generally defined as ratios of cubic combinations of the squared distances. When a particular pair of indices coincide, e.g. $m=i$, then they reduce to ratios of quadratic combinations. With a slight abuse of terminology we will call them as quadratic and cubic cross-ratios.}
\be
\label{eta} 
\left(\eta_k \right)_{ml}^{ij} \equiv \left(\eta_k \right)_{m,l}^{i,j} = \frac{Z_k^{ij}}{Z_k^{ml}} = \frac{X_{ij} X_{mk} X_{lk}}{X_{ik} X_{jk} X_{ml}} \,.
\ee
The splitting \eqref{intripleprod} is the consequence of the conformal covariance of \eqref{indef}: the leg-factor encodes the conformal transformation properties, while the function $\cI_n^{{\bm a}}({\bm \eta})$, called  a {\it bare conformal integral}, depends only  on the cross-ratios \eqref{eta}. The bare conformal integral is represented as the following $n(n-3)/2$-folded Mellin-Barnes integral\footnote{Appendix  \bref{app:functions} illustrates the formalism of Mellin-Barnes integrals using  examples of hypergeometric-type functions.}
\be
\begin{split}
\label{inbaremb}
\cI_n^{{\bm a}}({\bm \eta}) &= N_n^{\bm a} 
\intbrom \prod_{1\leq i < j \leq n-2} \left( \widehat{{\rm d}s}_{ij} \left((\eta_n)_{n-2,n-1}^{i,j}    \right)^{s_{ij}} \right) \prod_{k=1}^{n-3} \left(\widehat{{\rm d}t}_{k} \left( (\eta_n)_{n-2,n-1}^{k,n-1} \right)^{t_k} \right)  \\
&\times  \Gamma\Bigg(|\bm t_{1,n-3}| + \sum_{1\leq i < j \leq n-2} s_{ij} + a'_{n}\Bigg)  \prod_{l=1}^{n-3}\Gamma\Bigg(a_l+t_l +\sum_{j=1}^{l-1} s_{jl}+ \sum_{j=l+1}^{n-2} s_{lj}\Bigg) \\
&\times  \Gamma\Bigg(\alpha_{n-2,n}-\sum_{1\leq i < j \leq n-3} s_{ij} - |\bm t_{1,n-3}| \Bigg) \, \Gamma\Bigg(\alpha_{n-1,n} - \sum_{1\leq i < j \leq n-2} s_{ij}\Bigg)\,.
\end{split}
\ee
Here, $\alpha_{i,j}$ is defined in \eqref{notations_params} and the number of cross-ratios ${\bm \eta}$ equals the number of integrals on the right-hand side of \eqref{inbaremb}. By construction, the integrals over each variable are {\it balanced} (see e.g. \cite{Dubovyk:2022obc}), which means that the contours can be closed either to the left or to the right. In fact, the choice of a contour is determined by a domain of convergence in which one wants to obtain the asymptotic expansion of the conformal integral. In its turn, this determines a specific set of  cross-ratios ${\bm \eta}$.

As $n$ increases, the evaluation of the integral \eqref{inbaremb} quickly becomes difficult due to the complex pole structure of the integrand. In fact, the only well-understood example is given by the box ($n=4$) conformal integral which has been calculated by many authors in various contexts, see e.g. \cite{Dolan:2000uw} and references therein. Recently, however, significant progress has been made in the computation of higher-folded Mellin-Barnes integrals involving various computer algebra  methods \cite{Ananthanarayan:2020fhl,  Banik:2024ann}, in particular, the hexagon ($n=6$) conformal integral was calculated in \cite{Ananthanarayan:2020ncn, Banik:2023rrz}. 

We note finally that for some applications it is not necessary to know the whole integral \eqref{inbaremb}, but only that part of it which can be obtained by evaluating over a particular  subset of poles. E.g. this happens when calculating conformal blocks within the shadow formalism of CFT$_D$ \cite{SimmonsDuffin:2012uy, Rosenhaus:2018zqn,  Alkalaev:2024jxh}.

\subsection{Bipartite  Mellin-Barnes representation}
\label{sec:MB}

In what follows, we  suggest  a different way of representing conformal integrals \eqref{indef} in terms of Mellin-Barnes integrals which does not use the Symanzik conformal trick. The advantage of our approach is that a part of the $n$-point conformal integral can be computed effortlessly, while the remaining part is conjectured to be recovered by using  permutation  invariance of the whole conformal integral \eqref{indef} (see section \bref{sec:perm}). 

The first step is to remove  $1/|\bm \lambda_{1,n}|$  from the exponential in \eqref{inlambdas} by changing integration  variables  $\lambda_i \to |\bm \lambda_{1,n}| \lambda_i$ and obtaining 
\be
I_n^{{\bm a}}({\bm x}) = 2 N_n^{\bm a} \int_0^{+\infty} \prod_{i=1}^{n} \bigg( \frac{{\rm d} \lambda_i}{\lambda_i} \lambda_i^{a_i}  \bigg) \frac{1}{|\bm \lambda_{1,n}|^{D-|\bm a_{1,n}|}} \, \exp\bigg( - \sum_{1\leq i < j \leq n} \lambda_i \lambda_j X_{ij} \bigg)\,.
\ee
Then, the contribution of $|\bm \lambda_{1,n}|$ cancels  due to the conformality  constraint   \eqref{confcond}. Evaluating the integral over $\lambda_n$ yields  
\be
I_n^{{\bm a}}({\bm x}) = 2 N_{n-1}^{\bm a} \int_0^{+\infty} \prod_{i=1}^{n-1} \bigg( \frac{{\rm d} \lambda_i}{\lambda_i} \lambda_i^{a_i}  \bigg) \dps\bigg( \sum_{i=1}^{n-1} \lambda_i X_{i n} \bigg)^{-a_n} \, \exp\bigg( - \sum_{1\leq i < j \leq n-1} \lambda_i \lambda_j X_{ij} \bigg)\,.
\ee
Now, changing integration variables as $\lambda_i \to X_{i n}^{-1}\lambda_i$ (no summation over $i$) one finds  
\be
I_n^{{\bm a}}({\bm x}) = 2 N_{n-1}^{\bm a} \prod_{i=1}^{n-1} X_{i,n}^{-a_i} \int_0^{+\infty} \prod_{i=1}^{n-1} \bigg(\frac{{\rm d} \lambda_i}{\lambda_i} \lambda_i^{a_i} \bigg) \frac{1}{|\bm \lambda_{1,n-1}|^{a_n}} \, \exp\bigg( - \sum_{1\leq i < j \leq n-1} \lambda_i \lambda_j Z_n^{ij} \bigg)\,,
\ee
where $Z_n^{ij}$ are given by \eqref{zdef}. Rescaling $\lambda_i \to \lambda_i\sqrt{\lambda_{n-1}} $ for $i=1\,,..., n-2$ and $\lambda_{n-1} \to \sqrt{\lambda_{n-1}}$  one can evaluate the integral over $\lambda_{n-1}$: 
\be
\ba{l}
\dps
I_n^{{\bm a}}({\bm x}) = N_{n-1}^{\bm a}\, \Gamma(a'_n)  \prod_{i=1}^{n-1}X_{in}^{-a_i} \int_0^{+\infty} \prod_{j=1}^{n-2} \bigg( \frac{{\rm d} \lambda_j}{\lambda_j} \lambda_j^{a_j}  \bigg)
\vspace{2mm}
\\
\dps 
\hspace{50mm}\times \frac{\Big(1+|\bm \lambda_{1,n-2}|\Big)^{-a_n}} {\dps\bigg(\sum_{1\leq l < m \leq n-2} \lambda_l \lambda_m Z_n^{lm} + \sum_{k=1}^{n-2} \lambda_k Z_n^{k,n-1} \bigg)^{a'_n}}\,.
\ea
\ee
Introducing new variables by  $\lambda_i = s \sigma_i$ for $i=1,..., n-3$ and $\lambda_{n-2} = s (1 - |\bm \sigma| )$, where $|\bm \sigma| \equiv  |\bm \sigma_{1,n-3}|$, the integral over $s$ is recognized  as the integral representation of the Gauss hypergeometric function \eqref{2f1intinfty}. As a result, the conformal integral is again represented as the product \eqref{intripleprod} with the same leg-factor $L_n^{{\bm a}}({\bm x})$ \eqref{inleg}, but with the bare integral $\cI_n^{{\bm a}}({\bm \eta})$  written in the form
\be
\begin{split}
\label{inbare2f1}
\cI_n^{{\bm a}}({\bm \eta}) &= N_{n-2}^{\bm a}\, \Gamma \Bigg[
\begin{array}{l l}
a'_{n-1}, a'_n\\  \; \quad \hd
\end{array}
\Bigg]  \int_{\Omega} {\rm d}^{n-3} \sigma \; \frac{\dps\prod_{i=1}^{n-3} \sigma_i^{a_i-1} (1 - |\bm \sigma| )^{a_{n-2}-1}}{\dps\Big(1 - \bm \sigma \cdot \bm \xi \Big)^{a'_n}} \; {}_2 F_1 \Bigg[
\begin{array}{l l}
a'_{n-1} ,  a'_n \\
\; \quad \hd
\end{array}\bigg| 
1-\xi({\bm \sigma})
\Bigg]\,,
\end{split}
\ee
where the integration domain is the standard orthogonal simplex $\Omega = \{{\bm \sigma} \in \mathbb{R}^{n-3}: \sigma_j \geq 0\,, \, j=1,...,n-3\,;\, |{\bm \sigma}|\leq 1 \}$, and 
\be
\label{xin}
\ba{c}
\dps
\xi({\bm \sigma}) = \frac{\dps\sum_{1\leq i < j \leq n-3} \sigma_i \sigma_j (\eta_n)_{n-2,n-1}^{ij} + (1 - |\bm \sigma| ) \sum_{l=1}^{n-3} \sigma_l\, (\eta_n)_{n-2,n-1}^{l,n-2}}{\dps 1 - \bm \sigma \cdot \bm \xi }\,,
\vspace{2mm} 
\\
\dps
\bm \sigma \cdot \bm \xi = \sum_{l=1}^{n-3} \sigma_l \xi_l\,, \qquad \xi_i = 1 - (\eta_n)_{n-2,n-1}^{i,n-1}\,, \quad 1\leq i \leq n-3 \,,
\ea
\ee
where $\left(\eta_n \right)_{n-2,n-1}^{ij}$ are the cross-ratios \eqref{eta}.   
Note that there are $(n-3)(n-4)/2+(n-3) = (n-3)(n-2)/2$ terms in the numerator of $\xi({\bm \sigma})$,  each of which contains one cross-ratio. Adding  $(n-3)$ cross-ratios from the denominator one concludes that the bare conformal integral \eqref{inbare2f1} does depend on $n(n-3)/2$ cross-ratios of the type \eqref{eta}, as  discussed below \eqref{inbaremb}.\footnote{For $n>D+2$ the cross-ratios become dependent. Nonetheless,  we will assume that a complete set of cross-ratios for $n$ points in $\RR^{D}$ for any $D$ has   $n(n-3)/2$ elements.}

Let us now  express the bare integral \eqref{inbare2f1} through  the Mellin-Barnes integrals. It turns out to be convenient to use the analytic continuation formula for the Gauss hypergeometric function \eqref{2f1analyt} that results in splitting the bare integral into two terms
\be
\label{inbare}
\cI_n^{{\bm a}}({\bm \eta}) = \cI_n^{(1),{\bm a}}({\bm \eta}) + \cI_n^{(2),{\bm a}}({\bm \eta})\,,
\ee
where the terms on the right-hand side which we refer to as the {\it first and second bare integrals} are given by 
\be
\begin{split}
\label{in1}
\cI_n^{(1),{\bm a}}({\bm \eta}) &= N_n^{\bm a}\; \Gamma (
\alpha_{n-1,n}\,, a'_{n-1}\,, a'_n) \, \int_{\Omega} {\rm d}^{n-3} \sigma \; \frac{\dps\prod_{i=1}^{n-3}\sigma_i^{a_i-1} (1- |\bm \sigma|)^{a_{n-2}-1}}{(1 - \bm \sigma \cdot \bm \xi )^{a'_n}} \\
&\times {}_2 F_1 \Bigg[
\begin{array}{l l}
a'_{n-1} ,  a'_n \\
1-\alpha_{n-1,n}
\end{array}\bigg| 
\xi({\bm \sigma})
\Bigg],
\end{split}
\ee 
\be
\begin{split}
\label{in2}
\cI_n^{(2),{\bm a}}({\bm \eta}) &=  N_n^{\bm a}\; \Gamma (-
\alpha_{n-1,n}\,, a_{n-1}\,, a_n) \, \int_{\Omega} {\rm d}^{n-3} \sigma \; \frac{\dps\prod_{i=1}^{n-3}\sigma_i^{a_i-1} (1- |\bm \sigma|)^{a_{n-2}-1}}{(1 - \bm \sigma \cdot \bm \xi )^{a'_n}} \\
&\times \big(\xi({\bm \sigma}) \big)^{\alpha_{n-1,n}} \, {}_2 F_1 \Bigg[
\begin{array}{l l}
a_{n-1} ,  a_n \\
1+\alpha_{n-1,n}
\end{array}\bigg| 
\xi({\bm \sigma})
\Bigg].
\end{split}
\ee
This is a {\it bipartite representation} of the $n$-point (bare) conformal integral. A few  comments are in order. 

\begin{itemize}

\item Using a certain analytic continuation formula is a formal trick because  we cannot precisely describe  a domain of the functional argument \eqref{xin}. Nonetheless, this procedure will eventually lead to representing the conformal integral as a multivariate  powers series in $n(n-3)/2$ cross-ratios which is  valid for a particular arrangement of points $x_i$.

\item We note that one could directly represent the Gauss hypergeometric function in \eqref{inbare2f1} by means of  one of its Mellin-Barnes representations \eqref{2f1mb_2}. This yields a representation of the conformal integral in terms of the  Mellin-Barnes integrals given in appendix  \bref{app:MB_more}. However, this particular representation turns out to be less computationally efficient than that one we  derive in this section.

\item Similarly,  another  analytic continuation of the Gauss hypergeometric function could be used that would change the form of two terms in  \eqref{inbare}. Nonetheless, the convenience of choosing this particular partitioning of the conformal integral can be seen, in particular, in the  $n=3$ case, when the second bare integral  vanishes since the numerator of \eqref{xin} has no terms, while the first bare integral   leads to the star-triangle relation \cite{Symanzik:1972wj}
\be
\label{startriangle}
I_3^{{\bm a}}({\bm x}) = \rS_3^{\triple{123}}(\bm a)\, X_{12}^{-a'_3} X_{13}^{-a'_2} X_{23}^{-a'_1}\,, 
\qquad
\text{where}
\qquad
 \rS_3^{\triple{123}}(\bm a) \equiv  
\Gamma \Bigg[ 
\begin{array}{l l}
a'_1\,, a'_2\,, a'_3  \\
a_1, a_2, a_{3}
\end{array}
\Bigg],
\ee
which is commonly represented as the following diagrammatic equality 
\begin{figure}[h]
  \centering
  \[
  \begin{minipage}[h]{0.15\linewidth}
	\vspace{0pt}
	\includegraphics[width=\linewidth]{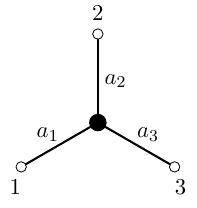}
   \end{minipage}
   \quad
    = \quad \rS_3^{\triple{123}}(\bm a) \quad
  \begin{minipage}[h]{0.15\linewidth}
	\vspace{0pt}
	\includegraphics[width=\linewidth]{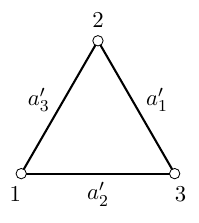}
   \end{minipage}
  \]
\end{figure}

\item In what follows we restrict $n\geq 3$ so that the $1$-point and $2$-point  conformal integrals fall out of our consideration. These lower-point integrals diverge and   require careful regularizations which have been extensively studied, see e.g. \cite{Gorishnii:1984te, Isaev:2003tk, Derkachov:2022ytx, Derkachov:2023xqq}. Notably, the 2-point conformal integral being proportional to $\delta(x_1-x_2)$ is the cornerstone of the shadow formalism of CFT$_D$  \cite{Ferrara:1972uq,Fradkin:1996is}. 

\end{itemize}

\paragraph{The term $\cI_n^{(1),{\bm a}}$.} Let us now explicitly calculate the first bare integral  \eqref{in1}. To integrate  over $\sigma_i$ we use the Mellin-Barnes representation for the Gauss hypergeometric function \eqref{2f1mb}:
\be
{}_2 F_1 \Bigg[
\begin{array}{l l}
\; a'_{n-1} ,  a'_n \\
1-\alpha_{n-1,n}
\end{array}\bigg| 
\xi({\bm \sigma})
\Bigg]  = \Gamma \Bigg[ 
\begin{array}{l l}
1-\alpha_{n-1,n}\\
\; a'_{n-1},  a'_{n}
\end{array}
\Bigg] \int_{-i\infty}^{+i\infty} \widehat{{\rm d} t} \, \Gamma \Bigg[ 
\begin{array}{l l}
a'_{n-1}+t, a'_n+t\\
\; 1-\alpha_{n-1,n}+t
\end{array}
\Bigg] \big(-\xi(\bm \sigma)\big)^t \,.
\ee
When $n=4$, the numerator of \eqref{xin} contains only one term, and thus we are left with the 1-dimensional integral over $\sigma_1$ which can be transformed into the Gauss hypergeometric function \eqref{2f1intinfty}. At  $n>4$ the functional argument \eqref{xin} is more complicated and its numerator can be converted into the $(n-3)(n-2)/2-1$ Mellin-Barnes integrals by means of the Mellin-Barnes expansion \eqref{mbexpansion}: 
\be
\label{xinmb}
\ba{l}
\dps
\bigg( \sum_{1\leq i < j \leq n-3} \sigma_i \sigma_j \, (\eta_n)_{n-2,n-1}^{ij} + (1-|\bm \sigma|) \sum_{l=1}^{n-3} \sigma_l \,(\eta_n)_{n-2,n-1}^{l,n-2} \bigg)^{t} = \frac{1}{\Gamma(-t)} 
\vspace{2mm} 
\\
\dps
\times \int_{-i \infty}^{+ i \infty} \prod_{l=1}^{n-4} \widehat{{\rm d}t}_l  
\int_{-i \infty}^{+ i \infty} \prod_{1\leq k < m \leq n-3} \widehat{{\rm d}s}_{km}\;  \Gamma\bigg(|\bm t_{1,n-4}| + \sum_{1\leq i < j \leq n-3} s_{ij} - t \bigg) 
\vspace{2mm} 
\\
\dps
\times \prod_{l=1}^{n-4}  \left((1-|\bm \sigma|) \sigma_l (\eta_n)_{n-2,n-1}^{l,n-2} \right)^{t_l} \prod_{1\leq i < j \leq n-3} \left(\sigma_i \sigma_j (\eta_n)_{n-2,n-1}^{ij}  \right)^{s_{ij}} 
\vspace{3mm} 
\\
\dps
\times \left((1-|\bm \sigma|) \sigma_{n-3} (\eta_n)_{n-2,n-1}^{n-3,n-2} \right)^{t-|\bm t_{1,n-4}| - \sum_{1\leq i < j \leq n-3} s_{ij}} \,.
\ea
\ee
The integral over $\sigma_i$ in \eqref{in1} is recognized as the integral representation of the Lauricella function $F_{D}^{(n-3)}$ \eqref{fdint}. Thus, we obtain the first bare integral  \eqref{in1} in the following form
\be
\label{in1mb}
\ba{l}
\dps
\hspace{0mm}
\cI_n^{(1),{\bm a}} = N_n^{\bm a}\, \Gamma \left(
1-\alpha_{n-1,n} \,, \alpha_{n-1,n}  \right) \, \int_{-i\infty}^{+i\infty} \frac{{\rm d} t}{2 \pi i}\, \Gamma \Bigg[ 
\begin{array}{l l}
a'_{n-1}+t\,, a'_n+t \\
1-\alpha_{n-1,n}+t
\end{array}
\Bigg]  \left(-(\eta_n)_{n-2,n-1}^{n-3,n-2} \right)^{t} 
\vspace{2mm} 
\\
\dps
\hspace{1mm} 
\times \intbrom \prod_{l=1}^{n-4} \left(\widehat{{\rm d}t}_l \,   \left((\eta_n)_{n-3,n-2}^{l,n-2}\right)^{t_{l}} \right)  
\intbrom \prod_{1\leq i < j \leq n-3} \left( \widehat{{\rm d}s}_{ij} \,   \left((\eta_n)_{n-3,n-2}^{ij}\right)^{s_{ij}} \right)
\vspace{2mm} 
\\
\dps
\hspace{1mm} 
\times\, \Gamma\Bigg(|\bm t_{1,n-4}| + \sum_{1\leq i < j \leq n-3} s_{ij} - t \Bigg) \, \Gamma \Bigg[ 
\begin{array}{l l}
C^{(1)}-|\bm B^{(1)}|, \bm B^{(1)} \\
\qquad C^{(1)}
\end{array}
\Bigg]\, F_{D}^{(n-3)} \Bigg[ 
\begin{array}{l l}
A^{(1)},\bm B^{(1)} \\
\quad C^{(1)}
\end{array}\bigg|
\bm \xi
\Bigg]\,,
\ea
\ee
where $\bm \xi = \{\xi_1,...,\xi_{n-3} \}$ is given in \eqref{xin}, and  the parameters, including $\Gamma(\bm B^{(1)}) \equiv  \Gamma(B_1^{(1)},...,B_{n-3}^{(1)})$, are expressed  in terms of  the propagator powers $a_i$ as 
\be
\label{in1laurparam}
\ba{c}
\dps
A^{(1)} = a'_n + t\,, 
\qquad 
C^{(1)} = |\bm a_{1,n-2}| + 2t \,, 
\qquad 
B_{n-3}^{(1)} = a_{n-3} + t - |\bm t_{1,n-4}| - \sum_{1\leq i < j \leq n-4} s_{ij} \,,
\vspace{2mm}
\\
\dps
B_l^{(1)} = a_l + t_l + \sum_{j=1}^{l-1} s_{jl} + \sum_{j=l+1}^{n-3} s_{lj}\,, \qquad  l=1,..., n-4  \,.
\ea
\ee
By construction, all the contours in \eqref{in1mb} for all integrals can be closed both to the left and right. When the contour over $t$ is closed to the right, only the poles coming from  $\Gamma\left(|\bm t_{1,n-4}| + \sum_{1\leq i < j \leq n-3} s_{ij} - t \right)$ contribute to the integral (see the third line in \eqref{in1mb}).  Calculating the integral over $t$ then leads to  that closing contours to the right in all the remaining integrals over $t_i$ and $s_{ij}$  only one set of poles contributes for each integration variable, namely, these are the poles coming from $\Gamma(-t_i)$ and $\Gamma(-s_{ij})$ (cf. the integration measure  \eqref{mb_measure}). In this way, the whole integral \eqref{in1mb} can be explicitly calculated as   
\be
\begin{split}
\label{in1sums}
\cI_n^{(1),{\bm a}} &= N_n^{\bm a}\, \Gamma \left(
1-\alpha_{n-1,n} \,, \alpha_{n-1,n} \right) \,\prod_{l=1}^{n-3} v_l^{a_l} \sum_{\{ k_l, p_{ij}\}=0}^{\infty}  \prod_{l=1}^{n-3}\frac{u_l^{k_l}}{k_l!} \prod_{1\leq i < j \leq n-3} \frac{w_{ij}^{p_{ij}}}{p_{ij}!}\\
&\times \Gamma \Bigg[
\begin{array}{l l}
a_{n-2}+ |\bm k|, a'_n+|\bm k| + |\bm p|, \cA^{(1)}, \bm \cB^{(1)} \\
1-\alpha_{n-1,n}+|\bm k| + |\bm p|, \cC^{(1)}
\end{array}
\Bigg] 
\,F_{D}^{(n-3)} \Bigg[ 
\begin{array}{l l}
\cA^{(1)},\bm \cB^{(1)} \\
\quad \cC^{(1)}
\end{array}\bigg|
\bm 1 -\bm v
\Bigg] \,,
\end{split}
\ee
where we used  the transformation formula for the Lauricella function \eqref{laurpfaff}; the parameters  are related to the propagator powers $a_i$ and summation indices as  
\be
\cA^{(1)} = a'_{n-1}+|\bm k|+ |\bm p|\,, \quad \cB_l^{(1)} = a_l + k_l + 2 p_l\,, \quad \cC^{(1)} = |\bm a_{1,n-2}| + 2(|\bm k|+|\bm p|)\,,
\ee
where $l=1,...,n-3$, and we set  
\be
\label{not1}
 |\bm p| = \sum_{1\leq i < j \leq n-3} p_{ij} = \sum_{l=1}^{n-3}p_l \,, \quad p_l = \frac{1}{2}\bigg(\sum_{j=1}^{l-1}p_{jl} + \sum_{j=l+1}^{n-3} p_{lj}\bigg)\,, \quad l=1,...,n-3\,.
\ee
The cross-ratios in \eqref{in1sums} are given by
\be
\label{incross}
\begin{split}
u_i &= (\eta_{n-1})_{n-2,n}^{i,n-2}\,, \qquad i=1,..., n-3\,, \\
w_{ij} &= (\eta_{n-1})_{n-2,n}^{i,j}\,, \qquad  1\leq i < j \leq n-3 \,, \\
v_i &= (\eta_{n-1})_{n-2,n}^{i,n}\,, \qquad  i=1,..., n-3 \,,
\end{split}
\ee
where $u_i, v_j$, and $w_{kl}$ are quadratic and cubic, respectively (see footnote \bref{fn:cross}).
Given the multivariate power series \eqref{in1sums} one can find its region of convergence, that, however, can be a rather complicated  problem. For simplicity, we suppose that such a series converges at least when the cross-ratios \eqref{incross} tend to specific points. In particular, since the Lauricella function $F_{D}$ \eqref{fd} is supported on the polydisk \eqref{polydisk}, we can claim that the multivariate  power series \eqref{in1sums} converges when
\be
\label{crossnope}
u_i,\, w_{kl} \to 0 
\quad
\text{and} 
\quad
v_j \to 1\,.
\ee 
Clearly, a range  of the cross-ratios is determined by a particular arrangement of  points $x_i$ and \eqref{crossnope}  corresponds to  the coincidence limit   
\be
x_n \to x_{n-1}\,.
\ee

\paragraph{The term $\cI_n^{(2),{\bm a}}$.} Consider the second bare integral  \eqref{in2}. Repeating  the same steps as in the previous paragraph one shows that
\be
\label{in2mb}
\ba{l}
\dps
\hspace{0mm}
\cI_n^{(2),{\bm a}} = N_n^{\bm a} \, \Gamma \left( 
1+\alpha_{n-1,n}, -\alpha_{n-1,n}
\right) 
\int_{-i\infty}^{+i\infty} \widehat{{\rm d} t}\, \Gamma \Bigg[ 
\begin{array}{l l}
a_{n-1}+t, a_n+t  \\
1+\alpha_{n-1,n}+t, -\alpha_{n-1,n}-t
\end{array}
\Bigg] \left( (\eta_n)_{n-2,n-1}^{n-3,n-2}\right)^{\alpha_{n-1,n} + t} 
\vspace{2mm}
\\
\dps
\times \intbrom \prod_{l=1}^{n-4} \left(\widehat{{\rm d}t}_l \left((\eta_n)_{n-3,n-2}^{l,n-2}\right)^{t_{l}} \right)  
\intbrom \prod_{1\leq i < j \leq n-3} \left( \widehat{{\rm d}s}_{ij} \left((\eta_n)_{n-3,n-2}^{i,j}\right)^{s_{ij}} \right) \\
\dps
\hspace{0mm}
\Gamma\Bigg(|\bm t_{1,n-4}| + \sum_{1\leq i < j \leq n-3} s_{ij} - t -\alpha_{n-1,n}\Bigg)   
\;\Gamma \Bigg[ 
\begin{array}{l l}
C^{(2)}-|\bm B^{(2)}|, \bm B^{(2)} \\
\quad C^{(2)}
\end{array}
\Bigg] 
\;F_{D}^{(n-3)} \Bigg[ 
\begin{array}{l l}
A^{(2)},\bm B^{(2)} \\
\quad C^{(2)}
\end{array}\bigg|
\bm \xi
\Bigg],
\ea
\ee
where 
\be
\begin{split}
\label{in2laurparam}
A^{(2)} &= a_{n-1} + t\,, \quad C^{(2)} = |\bm a_{1,n-2}| + 2 \alpha_{n-1,n} + 2t \,,  \\
B_l^{(2)} &= a_l + t_l +\sum_{j=1}^{l-1} s_{jl} + \sum_{j=l+1}^{n-3} s_{lj} \,, \quad l=1,..., n-4  \,, \\
B_{n-3}^{(2)} &= a_{n-3} + \alpha_{n-1,n}  + t  - |\bm  t_{1,n-4}| \;- \hspace{-3mm}\sum_{1\leq i < j \leq n-4} s_{ij}\,;
\end{split}
\ee
the arguments of the Lauricella function $F_{D}^{(n-3)}$ are given in \eqref{xin}. 

The pole structure of the second bare integral  $\cI_n^{(2),{\bm a}}({\bm x})$ is more complicated than that of the first bare integral  $\cI_n^{(1),{\bm a}}({\bm x})$ so we cannot immediately calculate the integrals in \eqref{in2mb}. The reason for this difference is the use of  particular analytic continuation formula for the Gauss hypergeometric function in the integral \eqref{inbare2f1}, which makes the two terms \eqref{in1} and \eqref{in2} different due to the additional factor of $\xi({\bm \sigma})$  in \eqref{in2}. In turn, when closing contours to the right, this gives an additional set of poles over $t$ in \eqref{in2mb}. This in turn produces different sets of cross-ratios and different analyticity domains. A priori one cannot say which contour should be chosen which will define the second bare integral $\cI_n^{(2),{\bm a}}({\bm x})$ in the  domain of the first integral \eqref{crossnope}. Nonetheless, the problem of  computing the second bare integral   can be addressed  by invoking invariance of the whole conformal integral  under the symmetric group.

\section{Constraints from the permutation invariance}
\label{sec:perm}

The symmetric group $\cS_n$ naturally  acts on  functions of coordinates and propagator powers by permuting elements  from $\bm x = (x_1,...,x_n)$ and $\bm a = (a_1, ...,a_n)$ as $x_i \to x_{\pi(i)}$ and $a_i \to a_{\pi(i)}$. The conformal integral \eqref{indef} is invariant under the  symmetric group:
\be
\label{in_perminv}
\forall \, \pi \in \cS_n: \qquad \mathrm{R}(\pi) \circ I_n^{{\bm a}}({\bm x}) = I_n^{{\bm a}}({\bm x})\,,
\ee
where $\mathrm{R}(\pi)$ stands for the representation of $\pi$.\footnote{We will use  the one-line notation for elements of the symmetric group. The identical permutation will be denoted as $e$ and $\mathrm{R}(e)=\id$.} This invariance property is manifestly seen on the conformal graph on fig. \bref{fig:vertex} which remains invariant with respect to any permutation of legs which are labelled by $(x_i, a_i)$. 

On the other hand, a splitting of the conformal integral into two terms provided by the bipartite representation is not invariant  with respect to  the symmetric group. In this way, having found the first term explicitly,  we expect to generate the unknown second term  by acting on the first one with particular permutations. 

\subsection{Analytic continuation}

 To simplify the study of permutation invariance, it is useful to analytically continue  the known first bare integral originally defined at ${\bm u}$, ${\bm w}\to 0$, ${\bm v}\to 1$ into the other domain, where all cross-ratios are near zero:
\be
\label{crossnope0}
{\bm u},\, {\bm w},\, {\bm v} \to 0\,. 
\ee 
Recalling the explicit expression   \eqref{in1sums} we  notice  that this can be achieved using  the analytic continuation formulas for the Lauricella function $F_D^{(n-3)}$ derived by Bezrodnykh in \cite{Bezrodnykh2016one, Bezrodnykh:2016two, Bezrodnykh2017, Bezrodnykh:2018acl, Bezrodnykh:2018review} (see also  appendix  \bref{app:functions}). Indeed,  applying  the analytic continuation formula \eqref{fdanalyt} one splits  \eqref{in1sums} into $n-2$ terms thereby obtaining the first bare integral in the domain \eqref{crossnope0} as the following sum\footnote{Here, the $q$-th term $\mathrm{U}_{n}^{(q),{\bm a}}$ arises from the $q$-th term $\mathrm{D}_q$  \eqref{fdanalytfunc}.}
\be
\label{in1analyt}
\cI_n^{(1),{\bm a}}({\bm \eta}) = \sum_{q=0}^{n-3} \mathrm{U}_{n}^{(q),{\bm a}}({\bm \eta})\,,
\ee
where the  $q=0$ term: 
\be
\label{varphi_n^0}
\ba{c}
\dps
\mathrm{U}_n^{(0),\bm a} =  \rS_n^{\triple{n-2,n-1,n}} \, \prod_{i=1}^{n-3} v_i^{a_i}\sum_{\{k_i, p_{ij}, m_i\}=0}^{\infty}  \prod_{i=1}^{n-3} \frac{u_i^{k_i} }{k_i!} \, \prod_{1 \leq i< j \leq n-3} \frac{w_{ij}^{p_{ij}} }{p_{ij}!}\;\; \prod_{l=1}^{n-3} \frac{\left( v_l \right)^{m_l} }{m_l!}
\vspace{2mm} 
\\
\dps
\hspace{30mm}\times \;\frac{\dps\prod_{l=1}^{n-3}(a_l)_{k_l+2 p_l+m_l} \, (a'_{n-1})_{|\bm k| + |\bm p| +|\bm m|} (-)^{|\bm p|}}{(1-\alpha_{n-1,n})_{|\bm k| + |\bm p| } \, (1-\alpha_{n-2,n-1})_{|\bm p| +|\bm m|}} \,,
\ea
\ee
the terms with $q=1,...,n-3$: 
$$
\mathrm{U}_n^{(q),\bm a} =\rS_n^{\triple{q,n-1,n}}\, v_q^{a'_n-|\bm a_{1,q-1}|} \prod_{l=1}^{q-1} v_l^{a_l} \sum_{\{ k_i,p_{ij},m_i\} =0}^{\infty} \prod_{l=1}^{n-3} \frac{(\cU_l^{(q)})^{k_l}}{k_l!} \prod_{1\leq i < j \leq n-3} \frac{(\cW_{ij}^{(q)})^{p_{ij}}}{p_{ij}!} \prod_{l=1}^{n-3} \frac{(\cV_l^{(q)})^{m_l}}{m_l!}  
$$
\be
\label{varphi_n^q}
\ba{l}
\dps
\times\; \frac{(a'_n - |\bm a_{1,q-1}|)_{|\bm m_q| + |\bm k_{q,n-3}|+|\bm p_{q}|} }{(1+a'_n - |\bm a_{1,q}|)_{|\bm m_q|  + |\bm k_{q+1,n-3}|+|\bm p_{q+1}|}}  (-)^{ |\bm k_{q+1,n-3}|+|\bm p_{q+1}| } \vspace{3mm} 
\\
\dps
\times\; \frac{\dps\prod_{l=1}^{q-1} (a_l)_{k_l+2 p_l+m_l} (a_{n-2})_{|\bm k|+m_q} \prod_{l=q+1}^{n-3} (a_l)_{k_l+2p_l+m_l}}{(1-\alpha_{n-1,n})_{|\bm k| + |\bm p|}}\,.
\ea
\ee
Here, $|\bm p|$ is defined in \eqref{not1}, other notations are given in appendix \bref{app:notations}. The new cross-ratios in \eqref{varphi_n^q} are expressed in terms of the previously introduced cross-ratios \eqref{incross} as follows
\be
\label{cross_UV_q}
\cU_l^{(q)} = \begin{cases}
               u_l\,, \quad l=1,...,q\,, \vspace{2mm}\\ 
               \dps \frac{u_l v_q}{v_l}\,, \quad l=q+1,...,n-3\,,
               \end{cases} \quad 
\cV_l^{(q)} =  \begin{cases}
			   \dps\frac{v_l}{v_q}\,, \quad l=1,..., q-1\,, \\ 
			   v_q\,,\quad l=q\,,\vspace{2mm}\\
			   \dps\frac{v_q}{v_l}\,, \quad l = q+1,...,n-3\,,
			   \end{cases}
\ee
\be
\label{cross_W_q}
\cW_{ij}^{(q)} = 
				\begin{cases} 
				\dps\frac{w_{ij}}{v_q}\,, \quad 1\leq i< j \leq q\,, \vspace{2mm}\\
				\dps\frac{w_{ij}}{v_j}\,, \quad i=1,...,q\,, \quad j = q+1,...,n-3\,,\vspace{2mm}\\
				\dps\frac{w_{ij} v_q}{v_i v_j}\,, \quad q+1 \leq i< j \leq n-3\,.
				\end{cases}
\ee
The overall coefficients in \eqref{varphi_n^0}-\eqref{varphi_n^q} are defined as 
\be
\label{star_n}
\rS_{n}^{\triple{ijk}}(\bm a) = \Gamma
\Bigg[
\begin{array}{l l}
|\bm a_{i,j}|-\hd \,, |\bm a_{j,k}|-\hd\,, |\bm a_{k,i}|-\hd \\
\qquad a_i\,, \qquad a_{j}\,, \qquad a_k
\end{array}
\Bigg], 
\qquad 
1\leq i<j<k \leq n\,.
\ee
These generalize the prefactor in the star-triangle relation \eqref{startriangle}. Note that the last multiplier in the numerator contains $|\bm a_{k,i}|$ with $k>i$, which is defined in \eqref{notations_params}.

It is important to stress that having the  expansion \eqref{in1analyt}  near the origin in the space of cross-ratios, i.e. at $u_i, w_{kl}, v_j \to  0$,  means that other arguments  must also tend  to zero, i.e.
\be
\label{UWV0}
\cU_i^{(q)},\; \cW_{kl}^{(q)},\; \cV_j^{(q)}  \to 0\,.
\ee
Since  $\cU_i^{(q)}, \cW_{kl}^{(q)}, \cV_j^{(q)}$ are (rational) functions of $u_i, w_{kl}, v_j$ \eqref{cross_UV_q}, then the condition \eqref{UWV0} determines a specific way  $u_i, w_{kl}, v_j$ tend to zero.\footnote{E.g. there are two  variables  such that $x,y\to 0$ and $x/y\to 0$ that  means that $x\to0$ faster than $y\to0$.} 

For future purposes,  we note  here that there are only two terms in the expansion  \eqref{in1analyt} which can be related by the action of a longest cycle $C_n \equiv (12...n) \in \cS_n$. Namely, 
\be
\label{varphi_cyclic_relation}
\cycle_n \circ \left(L_n^{\bm a}( \bm x)\, \mathrm{U}_n^{(0), \bm a}({\bm \eta}) \right) = L_n^{\bm a}( \bm x) \, \mathrm{U}_n^{(1), \bm a}({\bm \eta})\,,
\ee
where by $\cycle_n = \mathrm{R}(C_n)$ we denote the action of $C_n$ on functions of ${\bm x}$ and ${\bm a}$. 
The proof is fairly  straightforward if we  consider how $C_n$ changes both the leg-factor \eqref{inleg}:
\be
\cycle_n  \circ L_n^{\bm a}( \bm x) = v_1^{a'_n} u_1^{-a_{n-2}} \prod_{j=2}^{n-3}\left(\frac{w_{1j}}{v_j} \right)^{-a_j} \, L_n^{\bm a}( \bm x)\,, 
\ee 
and  the cross-ratios \eqref{incross}:
\be
\label{cyclingcrossr}
\ba{l}
\dps
\cycle_n \circ u_i = \frac{v_1}{v_{i+1}}\,, 
\qquad 
\cycle_n \circ u_{n-3} = v_1\,, 
\qquad 
\cycle_n \circ w_{ij} = \frac{w_{i+1, j+1} v_1}{v_{i+1} v_{j+1}}\,, 
\vspace{2mm}
\\
\dps
\cycle_n \circ v_i = \frac{w_{1,i+1}}{v_{i+1}}\,,  
\qquad 
\cycle_n \circ v_{n-3} = u_1\,,
\qquad
\cycle_n \circ w_{i,n-3}  = \frac{u_{i+1} v_1}{v_{i+1}}\,, 
\ea
\ee
where $i,j = 1,...,n-4$ and $i<j$. The relation \eqref{varphi_cyclic_relation} stems directly from the fact that both power series $\mathrm{U}_n^{(0), \bm a}$, $\mathrm{U}_n^{(1), \bm a}$ arise from the Lauricella function $F_D^{(n-3)}$ (see \eqref{gnjfd} and \eqref{fdanalytfunc}) by means of the  analytic continuation formula \eqref{fdanalyt}. Note that all other terms ($q\neq 0,1$) in \eqref{in1analyt} arise from the generalized hypergeometric series  $G^{(n-3,q)}$  (see \eqref{fdanalytfunc}) which are functions of different types for different $q$. Hence, other possible relations among $\mathrm{U}_n^{(q), \bm a}$ involving permutations different from the longest cycle cannot be obtained in such an obvious way. 

In the subsequent sections the bipartite representation \eqref{inbare} supplemented with the analytic continuation formula  \eqref{in1analyt} and  the symmetric group action is illustrated in detail on the particular examples of the box, pentagon, and hexagon conformal integrals. This will allow us to have a number of observations, conclusions, and exact results which will eventually be  summarized  in section \bref{sec:recon} in the form of the reconstruction conjecture.    

\subsection{Box integral}
\label{sec:box}

In this case, the first bare integral  $\cI_4^{(1),{\bm a}}(u,v)$ is supported on some domain on the $(u,v)$-plane of  cross-ratios 
\be
\label{cros4}
u= \frac{X_{12} X_{34} }{X_{13} X_{24}} \,, 
\qquad 
v = \frac{X_{14} X_{23} }{X_{13} X_{24}}\,.
\ee
Note that there are no  cubic cross-ratios in this case. According to the general formula \eqref{in1sums}, the integral can be  explicitly evaluated  as the asymptotic expansion at $(u,v) \to (0,1)$: 
\be
\begin{split}
\label{i41sums}
\cI_4^{(1),{\bm a}}(u,v) &= N_4^{\bm a} \, \Gamma \big(1-\alpha_{3,4},
\alpha_{3,4}\big)\, v^{a_1} \\
&\times \sum_{k=0}^{\infty}  \frac{u^{k}}{k!}\, 
\,\Gamma \Bigg[ 
\begin{array}{l l}
a_1+k\,, a_2+k\,, a'_3+k\,, a'_4+k \\
a_1+a_2+2k\,, 1-\alpha_{3,4}+k
\end{array}\Bigg] 
\,{}_2 F_1 \Bigg[
\begin{array}{l l}
a_1+k\,, a'_3+k \\
a_1+a_2+2k
\end{array}\bigg| 1-v
\Bigg]\,.
\end{split} 
\ee

We note that an arbitrary permutation from the symmetric group $\cS_4$    acting on the first bare integral  $\cI_4^{(1),\bm a}$  \eqref{i41sums} produces a  power series whose  domain of convergence is generally different from the original domain. For instance, a transposition $(23)\in \cS_4$ acts on the cross-ratios as
\be
\label{sig23}
\sigma_{23} \circ u = \frac{1}{u}\,, \qquad \sigma_{23} \circ v = \frac{v}{u}\,,
\ee
where $\sigma_{ij} \equiv  \mathrm{R}((ij))$ denotes the action of $(ij)$ on functions of ${\bm x}$ and ${\bm a}$. Obviously, such a permutation  sends the original domain $(u,v) \to (0,1)$ to $(u,v) \to \infty$. To return back to the original domain, the  action \eqref{sig23}  must be accompanied by making a suitable analytic continuation of the resulting power series for $\cI_4^{(1),\bm a}$. However, analytic continuation formulas are known only  for a limited number of generalized hypergeometric functions. Thus, to avoid using analytical continuation  we first try permutations which leave the cross-ratios invariant thereby keeping any domain of convergence intact.

\subsubsection{Kinematic group} 
\label{sec:kin}

Those elements of  the symmetric group $\cS_4$ which leave the cross-ratios invariant, 
\be
\exists\, \pi \in \cS_4: \qquad   \mathrm{R}(\pi) \circ u = u\,,
\quad
\mathrm{R}(\pi) \circ v = v\,,
\ee 
form a subgroup called {\it kinematic} \cite{Kravchuk:2016qvl}:  
\be
\label{kinematic_4}
\cS_4^{{\rm  kin}} = \Big\{e,\, (12)(34),\, (13)(24),\, (23)(14)\Big\} \equiv \mathbb{Z}_2 \times \mathbb{Z}_2 \subset \cS_4\,.
\ee 
By construction, a given element of the kinematic group  generates from the first bare integral  a power series in the same domain  $(u,v) \to (0,1)$. E.g. let us consider a permutation $(23)(14)$ and examine its action on $\cI_4^{(1), {\bm a}}(u,v)$. Note that despite  the cross-ratios are invariant such an action is not trivial since the permutation also interchanges $a_2 \leftrightarrow a_3$ and $a_1 \leftrightarrow a_4$.  To this end, one considers the invariance condition \eqref{in_perminv} which in this case takes the form  
\be
\label{i4perm}
\sigma_{23} \sigma_{14}  \circ I_4^{{\bm a}}({\bm x}) = I_4^{{\bm a}}({\bm x})\,.
\ee
Recalling the double product representation of the conformal integral \eqref{intripleprod} and using that the 4-point leg-factor \eqref{inleg}  transforms as
\be
\sigma_{23} \sigma_{14} \circ L_4^{{\bm a}}({\bm x}) = u^{\alpha_{3,4}} v^{a_1-a_4} L_4^{{\bm a}}({\bm x})\,,
\quad \text{where}\;\;
L_4^{{\bm a}}({\bm x}) =    X_{14}^{-a_1} X_{23}^{-a'_4} X_{24}^{-\alpha_{2,4}} X_{34}^{-\alpha_{3,4}} \,,
\ee
one obtains  that the bare conformal integrals are related as 
\be
\label{i4_perm_rel}
u^{\alpha_{3,4}} v^{a_1-a_4} \left( \sigma_{23} \sigma_{14} \circ \cI_4^{(1), {\bm a}}(u,v) + \sigma_{23} \sigma_{14} \circ \cI_4^{(2), {\bm a}}(u,v) \right) = \cI_4^{(1), {\bm a}}(u,v) +  \cI_4^{(2), {\bm a}}(u,v)\,.
\ee
It can be seen that the asymptotics of the first term  on the left-hand side at $(u,v) \to (0,1)$ is different from that of  $\cI_4^{(1), {\bm a}}$ \eqref{i41sums}. It is the prefactor $u^{\alpha_{3,4}} v^{a_1-a_4}$ which changes  the asymptotic behaviour  since by construction both $\cI_4^{(1), {\bm a}}$ and $\sigma_{23} \sigma_{14} \circ \cI_4^{(1), {\bm a}}$ have the same asymptotics in the considered domain. Thus, the first  term on the left-hand side must coincide with the second term on the right-hand side which is the second bare integral  $\cI_4^{(2), {\bm a}}$, i.e.
\be
\cI_4^{(2), {\bm a}}(u,v) = u^{\alpha_{3,4}} v^{a_1-a_4}\, \sigma_{23} \sigma_{14} \circ \cI_4^{(1), {\bm a}}(u,v)\,.
\ee
Explicitly, this gives  the second bare integral  as the  asymptotic expansion at $(u,v) \to (0,1)$:    
\be
\label{i42sums}
\ba{c}
\cI_4^{(2), {\bm a}}(u,v) =  N_4^{\bm a}\, \Gamma \big(1-\alpha_{1,2},
\alpha_{1,2}\big)\, u^{\alpha_{3,4}} v^{a_1} 
\vspace{2mm}
\\
\dps 
\times \, \sum_{k=0}^{\infty}  \frac{u^{k}}{k!}  \, \Gamma \Bigg[ 
\begin{array}{l l}
a'_1+k\,, a'_2+k\,, a_3+k\,, a_4+k \\
a_3+a_4+2k\,, 1-\alpha_{1,2}+k
\end{array}
\Bigg]\, {}_2 F_1 \Bigg[
\begin{array}{l l}
a'_2+k\,, a_4+k \\
a_3+a_4+2k
\end{array}\bigg| 1-v
\Bigg]\,,
\ea
\ee
cf. \eqref{i41sums}.  We conclude that the second bare integral  can be  reconstructed from the first one using a  certain permutation. On the other hand, going back to the Mellin-Barnes representation of $\cI_4^{(2), {\bm a}}$  \eqref{in2mb}, we note that in this case the integrals can be evaluated explicitly and the resulting function does coincide with \eqref{i42sums}.

The remaining permutations from $\cS_4^{{\rm  kin}}$ can be considered along the same lines. One  finds out that acting with $(13)(24)$ yields the same result, i.e. this permutation also reconstructs the second bare integral  $\cI_4^{(2), {\bm a}}$, while $(12)(34)$ keeps each bare integral invariant, i.e. 
\be
\sigma_{12} \sigma_{34}\circ \left(L_4^{\bm a}(\bm x)\, \cI_4^{(m), {\bm a}}(u,v)\right) = L_4^{\bm a}(\bm x)\, \cI_4^{(m), {\bm a}}(u,v)\,, \qquad m=1,2\,.
\ee
Thus, the conformal integral can be represented as 
\be
\label{4CI_kin}
\ba{c}
\dps
I_4^{\bm a}(\bm x) = \Big(\id + \sigma_{23} \sigma_{14}\Big) \circ L_4^{\bm a}(\bm x) \, \cI_4^{(1),\bm a}(\bm \eta) 
\vspace{2mm}
\\
\dps
\hspace{13mm} = \Big(\id + \sigma_{13} \sigma_{24}\Big) \circ L_4^{\bm a}(\bm x) \, \cI_4^{(1),\bm a}(\bm \eta)\,.
\ea
\ee
Here, a key concept is that the  conformal integral  in an appropriate coordinate domain can be generated by acting with particular permutations on a single function which will be called a {\it master} function.\footnote{This observation is known in the literature on (non-)conformal integrals. E.g. a similar permutation generated representation is valid for the non-parametric hexagon integral with three massive corners \cite{DelDuca:2011wh} or for the non-parametric pentagon integral \cite{Nandan:2013ip} (see also appendix  \bref{app:explicit_52}).}  

It should be noted that for $n>4$ the kinematic groups are all trivial, $\cS_n^{{\rm kin}} = \{e\}$  \cite{Kravchuk:2016qvl}. This motivates  the search for other subgroups of $\cS_n$ which also reconstruct the conformal integrals from a given set of master functions. For any $n$ the problem looks quite complicated.  Nonetheless, going back to the $n=4$ case and bearing in mind that then all the Mellin-Barnes integrals involved into the bipartite representation can be calculated explicitly,  an educated guess is to  consider the cyclic group $\mathbb{Z}_4\subset \cS_4$. In the higher-point case, this naturally generalizes to picking   $\mathbb{Z}_n \subset \cS_n$ as a {\it generating  group}: a subgroup of $\cS_n$ which allows reconstructing  the full conformal integral from a given set of master functions.

\subsubsection{Cyclic group} 
The cyclic group $\mathbb{Z}_4$ is
\be
\label{cyclic_4}
\mathbb{Z}_4 = \Big\{e,\, C_4, \, (C_4)^2,\, (C_4)^3\Big\} \subset \cS_4\,,
\ee 
where $C_4 = (1 2  3 4)$ is a longest cycle. The longest cycle permutes the two cross-ratios 
\be
\cycle_4 \circ u = v\,,
\quad
\cycle_4 \circ v = u\,,
\ee 
cf. \eqref{sig23}. Hence, the cycles interchange domains  $(u,v) \to (0,1)$ and  $(u,v) \to (1,0)$. Note, however, that $u,v=0$ is  {\it a fixed point} of $\mathbb{Z}_4$  that suggests that we first  analytically  continue the first bare integral  \eqref{i41sums} from $(u,v) \to (0,1)$ to  $(u,v) \to (0,0)$ and then consider cycles as generating permutations which reconstruct the full conformal integral near the origin on the $(u,v)$-plane. 

Thus,  we can apply the general analytic continuation formula \eqref{in1analyt}  which in this case takes the form 
\be
\label{i41_analyt}
\cI_4^{(1),{\bm a}}(u,v) = \mathrm{U}_4^{(0),{\bm a}}(u,v)+\mathrm{U}_4^{(1),{\bm a}}(u,v)\,,
\ee
where
\be
\label{varphi_01}
\begin{split}
\mathrm{U}_4^{(0),{\bm a}}(u,v) &= \rS_4^{\triple{234}}(\bm a) \, v^{a_1}\,  F_4 
\Bigg[
\begin{array}{l l}
a_1, a'_3 \\
1-\alpha_{3,4}, 1+\alpha_{1,4}
\end{array}\bigg| u,v 
\Bigg] \,, \\
\mathrm{U}_4^{(1),{\bm a}}(u,v) &= \rS_4^{\triple{134}}(\bm a) \, v^{a'_4} \,  F_4 
\Bigg[
\begin{array}{l l}
a_2, a'_4 \\
1-\alpha_{3,4}, 1-\alpha_{1,4}
\end{array}\bigg| u,v 
\Bigg]\,.
\end{split}
\ee
Here, $F_4$ is the fourth Appell function \eqref{f4} which resulted from using the splitting identity \eqref{f4split},  the coefficient $\rS_n^{\triple{ijk}}(\bm a)$ is defined in \eqref{star_n}.

Let us now consider  how $\mathbb{Z}_4$ acts on  $\mathrm{U}_4^{(0),{\bm a}}$ and $\mathrm{U}_4^{(1),{\bm a}}$. It turns out that these  two terms are related by the cyclic permutation as follows
\be
\label{cyclic_relation4_01}
\cycle_4 \circ \left(L_4^{\bm a}(\bm x)\, \mathrm{U}_4^{(0), \bm a}(u,v)\right) = L_4^{\bm a}(\bm x)\,  \mathrm{U}_4^{(1),{\bm a}}(u,v) \,,
\ee 
see the general statement \eqref{varphi_cyclic_relation}. Practically, this relation implies that there is only one {\it master} function which we choose to be $ L_4^{\bm a}(\bm x)\, \mathrm{U}_4^{(0),{\bm a}}(u,v)$. By construction, this new master function is supported near the origin  $(u,v)=(0,0)$. The cyclic group $\mathbb{Z}_4$ is of order four and, hence, its elements generate from the master  function three more functions which are also the fourth Appell functions supported on the same domain. Introducing the collective notation one can represent the resulting four functions (including the master one) as follows 
\be
\label{i4_terms}
\begin{split}
\BF_4^{\triple{234}}(\bm a|\bm x) &:= L_4^{\bm a}(\bm x)\, \mathrm{U}_4^{(0),{\bm a}}(u,v) =  \rS_4^{\triple{234}}(\bm a) \, \rV_4^{\triple{234}}(\bm a|\bm x)\, F_4 
\Bigg[
\begin{array}{l l}
a'_3, a_1 \\
1-\alpha_{3,4}, 1-\alpha_{2,3}
\end{array}\bigg| u,v  \Bigg] \,,  \\
\BF_4^{\triple{134}}(\bm a|\bm x)&:= (\cycle_4)^1\circ \BF_4^{\triple{234}} =  \rS_4^{\triple{134}}(\bm a) \, \rV_4^{\triple{134}}(\bm a|\bm x) \, F_4 
\Bigg[
\begin{array}{l l}
a_2, a'_4 \\
1-\alpha_{3,4}, 1-\alpha_{1,4}
\end{array}\bigg| u,v 
\Bigg], \\
\BF_4^{\triple{124}}(\bm a|\bm x) & :=  (\cycle_4)^2 \circ  \BF_4^{\triple{234}} =  \rS_4^{\triple{124}}(\bm a) \, \rV_4^{\triple{124}}(\bm a|\bm x) \, F_4 
\Bigg[
\begin{array}{l l}
a'_1, a_3 \\
1-\alpha_{1,2}, 1-\alpha_{1,4}
\end{array}\bigg| u,v 
\Bigg],\\
\BF_4^{\triple{123}}(\bm a|\bm x) & := (\cycle_4)^3\circ \BF_4^{\triple{234}}(\bm x) = \rS_4^{\triple{123}}(\bm a) \, \rV_4^{\triple{123}}(\bm a|\bm x)  \, F_4 
\Bigg[
\begin{array}{l l}
a'_2, a_4 \\
1-\alpha_{1,2}, 1-\alpha_{2,3}
\end{array}\bigg| u,v 
\Bigg],
\end{split}
\ee
where $\rS_n^{\triple{ijk}}(\bm a)$ are defined in \eqref{star_n}, the leg-factors $\rV_4^{\triple{ijk}}(\bm a|\bm x)$ are given by
\be
\label{i4_individual_leg}
\begin{split}
\rV_4^{\triple{234}}(\bm a|\bm x) &= X_{13}^{-a_1} X_{23}^{\alpha_{14}} X_{34}^{\alpha_{12}} X_{24}^{-a'_3} \,, 
\qquad 
\rV_4^{\triple{134}}(\bm a|\bm x) = X_{24}^{-a_2} X_{34}^{\alpha_{12}} X_{14}^{\alpha_{23}} X_{13}^{-a'_4} \,, 
\\
\rV_4^{\triple{124}}(\bm a|\bm x) &= X_{13}^{-a_3} X_{14}^{\alpha_{23}} X_{12}^{\alpha_{34}} X_{24}^{-a'_1} \,, 
\qquad 
\rV_4^{\triple{123}}(\bm a|\bm x) = X_{24}^{-a_4} X_{12}^{\alpha_{34}} X_{23}^{\alpha_{14}} X_{13}^{-a'_2}  \,, 
\end{split}
\ee
and  any two of them are related by a cyclic permutation, e.g.  $\rV_4^{\triple{134}} = \cycle_4 \circ \rV_4^{\triple{234}}$ (this also holds true for $\rS_4^{\triple{ijk}}$, e.g. $\rS_4^{\triple{134}} = \cycle_4 \circ \rS_4^{\triple{234}}$). The functions $\BF_4^{\triple{ijk}}(\bm a|\bm x)$ will be called {\it basis} functions. 

One can show that in terms of the basis functions both the first and second contributions to the conformal integral can be represented as  
\be
\label{i41_analyt1}
L_4^{\bm a}(\bm x)\, \cI_4^{(1),{\bm a}}(u,v) = \BF_4^{\triple{234}}(\bm a|\bm x) + \BF_4^{\triple{134}}(\bm a|\bm x)\,,
\ee
\be
L_4^{\bm a}(\bm x)\, \cI_4^{(2),\bm a}(u,v) = \BF_4^{\triple{124}}(\bm a|\bm x) + \BF_4^{\triple{123}}(\bm a|\bm x)\,.
\ee
Here, the first line is just \eqref{i41_analyt} in the new notation, while the second line results from analytically continuing the second bare integral  \eqref{i42sums} to the domain $(u,v) \to  0$ using the same continuation formula \eqref{2f1analyt}. In this way, one obtains yet another representation of the full conformal integral in terms of the generalized hypergeometric series generated from the master function  by the cyclic  group $\mathbb{Z}_4$:  
\be
\label{i4_result}
\ba{c}
\dps
I_4^{{\bm a}}({\bm x}) = \Big(\id + (\cycle_4)^1 +(\cycle_4)^2 + (\cycle_4)^3  \Big) \circ  \BF_4^{\triple{234}}(\bm a|\bm x) 
\vspace{2mm}
\\
\dps
= \BF_4^{\triple{234}}(\bm a|\bm x)  + \BF_4^{\triple{134}}(\bm a|\bm x)   + \BF_4^{\triple{124}}(\bm a|\bm x)   + \BF_4^{(123)}(\bm a|\bm x).
\ea
\ee
This expansion in basis functions is valid within the domain of convergence of the fourth Appell function $\sqrt{u} + \sqrt{v} < 1$ and reproduces the 4-point conformal integral in the form known in the literature.\footnote{The explicit calculation of the 4-point conformal integral has a long history which is discussed in detail in e.g. \cite{Dolan:2000uw}.} Finally, note that the kinematic group considered in the previous section also acts on the basis functions \eqref{i4_terms} so that their  sum \eqref{i4_result} remains invariant,  see appendix  \bref{app:kin}.

\subsubsection{Consistency checks}
\label{sec:box_checks}

There are at least two possible  non-trivial ways to test the resulting expressions.  First,  the most direct way to check the  $n$-point parametric conformal integral is to reduce a number of points by one, i.e. $n\to n-1$ along with $a_j\to 0 $ for some $j$ from 1 to $n$ and then constrain the remaining propagator powers by the $(n-1)$-point conformality  condition 
\be
\sum_{{i=1 \atop i\neq j}}^n a_i =D\,,
\ee 
cf. \eqref{confcond}. The result must be given by the $(n-1)$-point parametric conformal integral.    Second, one can consider the non-parametric case i.e. choose all $a_i=1$, and then compare the resulting expression with those  known in the literature which were 
obtained by other methods. These two checks are described in detail below.\footnote{One may also consider  conformal integrals as one of points goes to infinity, e.g. $x_{n} \to \infty$, that partially breaks conformal invariance. In this limit, the 4-point conformal integral is a particular case of the triangle integral evaluated by Boos and Davydychev \cite{Boos:1987bg, Boos:1990rg}. Note that conformal integrals with partially broken conformal invariance are instrumental  within the shadow formalism for CFT on  nontrivial backgrounds \cite{Alkalaev:2024jxh}.} 

\paragraph{Reduction to the star-triangle.} From the very definition of the 4-point integral  \eqref{indef} it follows that if one of propagator powers equals zero, e.g.  $a_4 = 0$, then the  4-point conformal integral reduces to the 3-point conformal integral:
\be
\label{i4limit1}
I_4^{\bm a}({\bm x})\Big|_{a_4=0} =\; I_3^{a_1,a_2,a_3}(x_1,x_2,x_3)\,.
\ee
Given the 4-point conformal integral calculated in the form  \eqref{i4_result} one can see from   \eqref{i4_terms} that by setting $a_4=0$ all basis functions  except $\BF_4^{\triple{123}}$ vanish due to $\Gamma(a_4\to 0) \to \pm \infty$ in their denominators. The only non-vanishing contribution then leads to\footnote{Note that the product $ \rV_4^{\triple{ijk}}\, \rS_4^{\triple{ijk}}$ can then be thought of as the 4-point  generalization of the right-hand side of the star-triangle relation \eqref{startriangle}.}
\be
\label{i4limit2}
I_4^{\bm a}({\bm x})\Big|_{a_4=0} = \; \Big(  \rS_4^{\triple{123}}(\bm a)\, \rV_4^{\triple{123}}  (\bm a|\bm x)  \Big)\Big|_{a_4=0}  \,,
\ee
where we used that the Appell function $F_4$ equals 1 if one of the upper parameters is 0, see  \eqref{f4}. Substituting  \eqref{star_n} and \eqref{i4_individual_leg} one finds out  that the left-hand side  of \eqref{i4limit2} reproduces the star-triangle relation \eqref{startriangle} provided that $a_1+a_2+a_3 = D$. The remaining limiting cases $a_i= 0$, $i=1,2,3$, can be analyzed along the same lines. The only difference  is which of the series \eqref{i4_terms} is non-zero in this limit. However, the numbering of basis functions is designed in such a way that when $a_l=0$ is imposed, a  basis function  $\BF_4^{\triple{ijk}}$ survives which does not have $l$ among  upper indices, i.e. $i,j,k \neq l$.

\paragraph{Non-parametric box.} For the 4-point conformal integral, the transition from the parametric to  non-parametric case was considered in \cite{Davydychev:1992eww, Dolan:2000uw}. Here, we reproduce a part of this analysis for completeness.  The subtlety is that when choosing $a_i=1$ in $D=4$ (cf. \eqref{confcond}) the resulting integral diverges because the basis functions  \eqref{i4_terms} contain divergent $\Gamma$-function prefactors. Introducing  a cutoff  parameter $\epsilon \to 0$ as
\be
\label{i4_param_epsilon}
a_1=a_2=a_3=1\,, \qquad a_4 = 1-2\epsilon \qquad \Rightarrow \qquad D=4-2\epsilon\,,
\ee
and expanding the 4-point conformal integral \eqref{i4_result} around $\epsilon = 0$ one finds
\be
\label{CI_BW}
I_4^{1,1,1,1-2\epsilon}({\bm x}) = \frac{1}{X_{13} X_{24}} \frac{ \Phi(u,v)}{(1-u-v)^2-4 u v} + O(\epsilon)\,,
\ee
where the Bloch-Wigner function $\Phi(u,v)$ is expressed in terms of polylogarithms \cite{Davydychev:1992eww}: 
\be
\label{blochuv}
\begin{split}
\Phi(u,v) &=  \frac{\pi^2}{3} + \ln u \ln v + \ln\frac{1+u-v-\lambda(u,v)}{2u}  \ln\frac{1-u+v-\lambda(u,v)}{2v}    \\
&+ 2\ln\frac{1+u-v-\lambda(u,v)}{2u}  + 2 \ln\frac{1-u+v-\lambda(u,v)}{2v}  \\
&- 2\Li2\frac{1+u-v-\lambda(u,v)}{2} - 2\Li2\frac{1-u+v-\lambda(u,v)}{2} \,,
\end{split}
\ee
where $\Li2$ is a dilogarithm and $\lambda(u,v) = \sqrt{(1-u-v)^2 - 4 u v}$. Introducing variables $u = z(1-y)$ and $v = y(1-z)$ and using the identity $\Li2 z + \Li2(1-z) = \pi^2/6-\ln z \ln(1-z)$ the Bloch-Wigner function \eqref{blochuv} can be cast into the form  \cite{Dolan:2000uw}
\be
\label{blochzy}
\Phi(z,y) =  \ln\big(y(1-z)\big) \ln\frac{z}{1-y} + 2 \Li2(1-z) -2 \Li2 y \,.
\ee
Then, the non-parametric box integral \eqref{CI_BW} can be given in a standard form \cite{Davydychev:1992eww, Dolan:2000uw} 
\be
\label{i41111}
I_4^{1,1,1,1}({\bm x}) = \frac{1}{X_{13} X_{24}} \frac{ \Phi(z,y)}{1-z-y}\,.
\ee

\subsubsection{Summary of the box integral reconstruction} 
\label{sec:sum}

Let us briefly formulate an emerging  strategy of reconstructing  a given conformal integral in some coordinate domain. We will keep $n$ arbitrary assuming that the box integral example can be  directly generalized to the higher-point case.

\begin{itemize}

\item Having calculated  the first bare integral in the form of multivariate hypergeometric series \eqref{in1sums}, we analytically continue the resulting function from the original region to another region around the origin in the space of cross-ratios $({\bm u}, {\bm w}, {\bm v}= 0)$ by means of the  analytic continuation formula \eqref{in1analyt} which splits $\cI_n^{(1),\bm a}$ into $n-2$ additive terms ${ \rm U}_n^{(q),\bm a}$, $q=0,1,...,n-3$. 

\item Among these $n-2$ terms we single out  $n-3$ terms which  are not related through the action of the cyclic group (see \eqref{varphi_cyclic_relation})
\be
\label{cyclic_n}
\mathbb{Z}_{n} = \{e, C_n, (C_n)^2, ..., (C_n)^{n-1}\} \subset \cS_n\,,
\ee
where $C_n = (123...n)\in \cS_n$ is a longest cycle.

\item Combining cyclically independent functions  with the leg-factor \eqref{inleg} one defines $n-3$ master functions 
\be
\label{master_q}
\begin{split}
\BF_n^{\triple{q,n-1,n}}(\bm a|\bm x) &= L_n^{\bm a}(\bm x)\, { \rm U}_n^{(q),\bm a}\,, \quad q=2,3,...,n-3\,, \\
\BF_n^{\triple{n-2,n-1,n}}(\bm a|\bm x) &= L_n^{\bm a}(\bm x)\, { \rm U}_n^{(0),\bm a}\,.
\end{split}
\ee

\item The idea of reconstruction: the conformal integral $I_n^{\bm a}(\bm x)$ is  supposed to be given by a sum of functions obtained by acting with all elements of $\mathbb{Z}_{n}$ on the master functions.  

\item The resulting  expression can be checked in at least two ways:
\begin{itemize}

\item The $n$-point conformal integral reduces to the $(n-1)$-point one upon setting one of propagator powers ${\bm a}$ equal to zero 
\be
\label{reduction}
\begin{split}
I_n^{a_1,...,a_{j-1},a_j,a_{j+1},...,a_n}&(x_1,...,x_{j-1},x_j,x_{j+1},...,x_n)\bigg|_{a_j=0} \\
&=I_{n-1}^{a_1,...,a_{j-1},a_{j+1},...,a_n}(x_1,...,x_{j-1},x_{j+1},...,x_n)\,,
\end{split}
\ee 
for any $j=1,...,n$. This reduction condition obviously follows from the very definition \eqref{indef}.

\item Going to the non-parametric case, the resulting  $I_n^{(1,1,...,1)}(\bm x)$ can be compared with expressions known  in the literature, if any.  

\end{itemize}

\end{itemize}

In the next sections \bref{sec:penta} and \bref{sec:hex} we  examine how this scheme works for the pentagon and hexagon integrals and then in section \bref{sec:recon} we will be able to formulate  the general reconstruction conjecture.

\subsection{Pentagon integral}
\label{sec:penta}

For $n=5$ the general formula \eqref{in1sums} represents the first bare integral as the asymptotic expansion at $u_{1,2}, w_{12} \to 0$, $v_{1,2} \to 1$:
\be
\begin{split}
\label{i51sums}
\cI_5^{(1),{\bm a}} &= N_5^{\bm a} \,
\Gamma \big( 
1-\alpha_{4,5} \,, \alpha_{4,5}  \big) \, v_1^{a_1}  v_2^{a_2}  \sum_{k_1, k_2, p_{12} =0}^{\infty}  \frac{u_1^{k_1}}{k_1!}\,
\frac{u_2^{k_2}}{k_2!}\,
\frac{w_{12}^{p_{12}}}{p_{12}!} \\
&\times  \Gamma \Bigg[ 
\begin{array}{l l}
a_1+k_1+p_{12}, a_2+k_2+p_{12}, a_3+|\bm k|, a'_4+|\bm k|+p_{12}, a'_5+|\bm k|+p_{12}  \\
\qquad \quad a_1+a_2+a_3+2(|\bm k|+p_{12}), \qquad  1-\alpha_{4,5}+|\bm k|+p_{12}
\end{array}
\Bigg]\\
&\times  F_1 \Bigg[
\begin{array}{l l}
\, a'_4+|\bm k|+p_{12}\,,  a_1+k_1+p_{12} \,, a_2+k_2+p_{12} \\
\qquad a_1+a_2+a_3+2(|\bm k|+p_{12})
\end{array}\bigg|\, 1 - v_1\,, 1 - v_2\,
\Bigg],
\end{split}
\ee
where $F_1$ is the first Appell function \eqref{f1}. Among the respective cross-ratios, one is cubic and the others are quadratic:
\be
\begin{split}
\label{i5cross}
u_1 &= \frac{X_{13}X_{45}}{X_{14}X_{35}} \,, \qquad  u_2 = \frac{X_{23}X_{45}}{X_{24}X_{35}}\,,  \qquad w_{12} = \frac{X_{12}X_{34}X_{45}}{X_{14} X_{24} X_{35}} \,, \\
v_1 &= \frac{X_{15} X_{34}}{X_{14}X_{35}}\,, \qquad v_2 = \frac{X_{25} X_{34}}{X_{24}X_{35}} \,.
\end{split}
\ee

According to the general strategy outlined in the previous section, we can try to reconstruct the conformal integral from a set of master functions which  can be found by particular analytic continuation of  the first bare integral. The analytic continuation  formula \eqref{in1analyt}-\eqref{varphi_n^q} splits the first bare integral \eqref{i51sums} into three terms
\be
\label{i51_analyt}
\cI_5^{(1),{\bm a}}(\bm \eta) = \mathrm{U}_5^{(0),{\bm a}}(\bm \eta)+ \mathrm{U}_5^{(1),{\bm a}}(\bm \eta) +  \mathrm{U}_5^{(2),{\bm a}}(\bm \eta)\,,
\ee
where 
\be
\label{varphi5_0}
\ba{r}
\dps
\mathrm{U}_5^{(0),{\bm a}}(\bm \eta) = \rS_5^{\triple{345}} \,v_1^{a_1}  v_2^{a_2}  \, \sum_{k_1,k_2,p_{12},m_1,m_2=0}^{\infty} 
\frac{u_1^{k_1}}{k_1!}
\frac{u_2^{k_2}}{k_2!}
\frac{w_{12}^{p_{12}}}{p_{12}!}
\frac{v_1^{m_1}}{m_1!}
\frac{v_2^{m_2}}{m_2!} 
\vspace{2mm}
\\
\dps
\times \frac{(a_1)_{k_1+p_{12}+m_1}\, (a_2)_{k_{2}+p_{12}+m_2}\, (a'_4)_{|\bm k|+p_{12}+|\bm m|} \,(-)^{p_{12}} }{(1-\alpha_{4,5})_{|\bm k|+p_{12}}\, (1-\alpha_{3,4})_{p_{12}+|\bm m|}} \;,
\ea
\ee
\be
\label{varphi5_1}
\ba{r}
\dps
\mathrm{U}_5^{(1),{\bm a}}(\bm \eta) = \rS_5^{\triple{145}}\, v_1^{a'_5}\,  \sum_{k_1,k_2,p_{12},m_1,m_2=0}^{\infty} 
\frac{u_1^{k_1}}{k_1!}
\frac{(u_2 v_1/v_2 )^{k_2}}{k_2!}
\frac{(w_{12}/v_2)^{p_{12}}}{p_{12}!}
\frac{v_1^{m_1}}{m_1!}
\frac{(v_1/v_2)^{m_2}}{m_2!} 
\vspace{2mm}
\\
\dps
\times \frac{(a_2)_{k_2+p_{12}+m_2} (a_3)_{|\bm k|+m_1} (a'_5)_{|\bm k|+p_{12}+|\bm m|}\, (-)^{k_2} }{(1-\alpha_{4,5})_{|\bm k|+p_{12}} (1-\alpha_{1,5})_{k_2+|\bm m|}} \,,
\ea
\ee 
\be
\label{varphi5_2}
\ba{r}
\dps
\mathrm{U}_5^{(2),{\bm a}}(\bm \eta) = \rS_5^{\triple{245}}\, v_1^{a_1} v_2^{-\alpha_{15}}\,  \sum_{k_1,k_2,p_{12},m_1,m_2=0}^{\infty} 
\frac{u_1^{k_1}}{k_1!}
\frac{u_2^{k_2}}{k_2!}
\frac{(w_{12}/v_2)^{p_{12}}}{p_{12}!}
\frac{(v_1/v_2)^{m_1}}{m_1!}
\frac{v_2^{m_2}}{m_2!} 
\vspace{2mm}
\\
\dps
\times \frac{(a_1)_{k_1+p_{12}+m_1} (a_3)_{|\bm k|+m_2}  (-\alpha_{1,5})_{m_2+k_2-m_1}\, (-)^{p_{12}} }{(1-\alpha_{4,5})_{|\bm k|+p_{12}} (1+\alpha_{3,4})_{m_2-m_1-p_{12}} }\,.
\ea
\ee
These functions are supported near the origin of coordinates  \eqref{crossnope0} such that the cross-ratios satisfy the condition \eqref{UWV0}.    

\subsubsection{Cyclic group}

The cyclic group is $\mathbb{Z}_5 = \big\{e,\, C_5, \, (C_5)^2,\, (C_5)^3, \, (C_5)^4\big\} \subset \cS_5$,  where $C_5 = (1 2 3 4 5)$ is a longest cycle. The general statement \eqref{varphi_cyclic_relation} about the relation of two terms  in the analytic continuation formula  for the first bare integral now reads as 
\be
\label{varphi5_01}
\cycle_5 \circ \left( L_5^{\bm a}(\bm x) \, \mathrm{U}_5^{(0),{\bm a}}(\bm \eta)\right) = L_5^{\bm a}(\bm x)\, \mathrm{U}_5^{(1),{\bm a}}(\bm \eta) \,,
\ee
where the 5-point leg-factor \eqref{inleg} is given by 
\be
\label{i5leg} 
L_5^{{\bm a}}({\bm x}) =  X_{15}^{-a_1} X_{25}^{-a_2} X_{34}^{-a'_5} X_{35}^{-\alpha_{3,5}} X_{45}^{-\alpha_{4,5}}\,.
\ee
It means that among three functions in \eqref{i51_analyt} there are two  which are not related by cyclic permutations, e.g. these are  $\mathrm{U}_5^{(0),{\bm a}}$ and $\mathrm{U}_5^{(2),{\bm a}}$. Then, adding the  leg-factor we introduce two master functions as
\be
\label{i5_master_functions}
\begin{split}
\BF_5^{\triple{345}}(\bm a|\bm x) & := L_5^{\bm a}(\bm x)\,  \mathrm{U}_5^{(0),{\bm a}}(\bm \eta) =  \rS_5^{\triple{345}}\,\rV_5^{\triple{345}}\, 
\mathrm{P}_1 \Bigg[ 
\begin{array}{l l}
a_1\,, a_2\,, a'_4 \\
1-\alpha_{45}, 1- \alpha_{34}\,,
\end{array}\Bigg| u_1, w_{12}, u_2, v_1, v_2
\Bigg]\,,\\
\BF_5^{\triple{245}}(\bm a|\bm x) & := L_5^{\bm a}(\bm x)\, \mathrm{U}_5^{(2),{\bm a}}(\bm \eta) =  \rS_5^{\triple{245}}\, \rV_5^{\triple{245}}\,   \mathrm{P}_2 \Bigg[ 
\begin{array}{l l}
a_1\,, a_3 \,, -\alpha_{15}\\
1- \alpha_{45}, 1+\alpha_{34}
\end{array}\Bigg| u_1, \frac{w_{12}}{v_2}, u_2, \frac{v_1}{v_2}, v_2
\Bigg] \,,
\end{split}
\ee
where we  defined two generalized hypergeometric functions $\mathrm{P}_1$  \eqref{pentagon_p1} and   $\mathrm{P}_2$ \eqref{pentagon_p2} as well as  two leg-factors 
\be
\label{i5_individual_leg}
\ba{l}
\dps
\rV_5^{\triple{345}}(\bm a|\bm x) = X_{14}^{-a_1} X_{24}^{-a_2} X_{45}^{-\alpha_{45}} X_{34}^{-\alpha_{34}} X_{35}^{-a'_4} 
\,, 
\vspace{2mm}
\\
\dps
\rV_5^{\triple{245}}(\bm a|\bm x) = X_{14}^{-a_1} X_{35}^{-a_3} X_{24}^{\alpha_{15}} X_{45}^{-\alpha_{45}}  X_{25}^{\alpha_{34}}\,,
\ea
\ee
cf. \eqref{i4_individual_leg}. 

The cyclic group $\mathbb{Z}_5$ is of order five so that  one can construct four more basis functions from each master function. In this way, we obtain  {\it ten} basis functions which are listed in appendix \bref{app:explicit_5}. Note that our notation for the basis functions is designed in such a way that the action of a cyclic permutation on the master function $\BF_5^{\triple{ijk}}$ cyclically permutes the upper indices $i,j,k$. In particular, it follows  that the two master functions \eqref{i5_master_functions} as well as  prefactors \eqref{i5_individual_leg} are not related to each other since there are no cyclic permutations which turn $345$ into $245$. Thus, the set of basis functions is split into two subsets of cyclically related functions, i.e. into two $\mathbb{Z}_5$-orbits.

Then, the 5-point parametric conformal integral is represented as the $\mathbb{Z}_5$-invariant sum of  ten basis functions:
\be
\label{i5_result}
\ba{c}
\dps
I_5^{{\bm a}}({\bm x}) = \sum_{j=0}^{4} (\cycle_5)^{j} \circ \Big( \BF_5^{\triple{345}}(\bm a|{\bm x}) + \BF_5^{\triple{245}}(\bm a|{\bm x})\Big)
\vspace{1mm}
\\
\dps
= \BF_5^{\triple{345}}(\bm a|{\bm x}) + \BF_5^{\triple{145}}(\bm a|{\bm x}) + \BF_5^{\triple{125}}(\bm a|{\bm x}) + \BF_5^{\triple{123}}(\bm a|{\bm x}) + \BF_5^{\triple{234}}(\bm a|{\bm x})
\vspace{1mm}
\\
\dps
\quad + \BF_5^{\triple{245}}(\bm a|{\bm x}) + \BF_5^{\triple{135}}(\bm a|{\bm x}) + \BF_5^{\triple{124}}(\bm a|{\bm x}) + \BF_5^{\triple{235}}(\bm a|{\bm x}) + \BF_5^{\triple{134}} (\bm a|{\bm x})
\,.
\ea
\ee
The expansion is defined in a domain near the origin in the space of cross-ratios. By construction, both sides here are invariant under cyclic permutations. 

Note that one can define a generalization of the kinematic group which we call an {\it extended} kinematic group. Given two master functions \eqref{i5_master_functions}, such a  group generates the same set of basis functions thereby providing an equivalent way of reconstructing  the conformal pentagon integral. Details are given in  appendix \bref{app:kin}.

\subsubsection{Reduction to the box integral}

In  appendix  \bref{app:explicit_52} we check that going to the non-parametric case we indeed   reproduce the formula for the non-parametric pentagon integral known in the literature \cite{Nandan:2013ip}. Of course, we can compare only in a given coordinate domain. Below we examine the reduction to the box integral.   

Setting one of propagator powers  equal to zero, e.g. $a_5 = 0$, the pentagon integral has to reduce to the box integral:
\be
\label{i5_limit_1}
I_5^{\bm a}({\bm x})\Big|_{a_5=0} =\; I_4^{a_1,a_2,a_3,a_4}(x_1,x_2,x_3,x_4)\,.
\ee
In fact, it is quite easy  to verify this relation provided that $I_5^{\bm a}({\bm x})$ is represented as  \eqref{i5_result}.  At $a_5 = 0$  only those basis functions $\BF_5^{\triple{ijk}}$ survive which upper indices $i,j,k \neq 5$, while the remaining basis functions vanish due to the presence of $\Gamma(a_5\to 0) \to \pm \infty$ in their  denominators, see \eqref{star_n}. As a result, 
\be
\label{i5_limit_2}
I_5^{\bm a}({\bm x})\Big|_{a_5=0} = \Big( \BF_5^{\triple{123}}(\bm a|{\bm x}) + \BF_5^{\triple{234}}(\bm a|{\bm x}) + \BF_5^{\triple{124}}(\bm a|{\bm x}) +  \BF_5^{\triple{134}}(\bm a|{\bm x})  \Big)\Big|_{a_5=0} \,.
\ee
The basis functions are written in terms of functions \eqref{pentagon_p1} and \eqref{pentagon_p2} which are reduced to the fourth Appell function \eqref{f4} when one of the upper parameters equals zero, i.e.
\be
\label{p1_p2_f4}
\mathrm{P}_1 
\Bigg[
\begin{array}{l l}
0, A, B \\
C_1, C_2
\end{array}\Bigg| \bm \xi
\Bigg] =
\mathrm{P}_2
\Bigg[
\begin{array}{l l}
0, A, B \\
C_1, C_2
\end{array}\Bigg| \bm \xi
\Bigg] 
=
F_4 
\Bigg[
\begin{array}{l l}
A, B \\
C_1, C_2
\end{array}\Bigg| \xi_3, \xi_5
\Bigg] \,.
\ee
On the other hand, the prefactors $\rS_5^{\triple{ijk}}(\bm a)\, \rV_5^{\triple{ijk}}(\bm a|\bm x)$ are reduced as 
\be 
\label{v5_reduction}
\begin{split}
\rS_5^{\triple{123}} \,\rV_5^{\triple{123}}  \Big|_{a_5=0} &= \rS_4^{\triple{123}}\,\rV_4^{\triple{123}}\,, 
\qquad 
\rS_5^{\triple{234}}\,\rV_5^{\triple{234}}  \Big|_{a_5=0}  = \rS_4^{\triple{234}}\,\rV_4^{\triple{234}}\,, 
\\
\rS_5^{\triple{124}}\,\rV_5^{\triple{124}} \Big|_{a_5=0} &= \rS_4^{\triple{124}}\,\rV_4^{\triple{124}}\,, 
\qquad 
\rS_5^{\triple{134}}\,\rV_5^{\triple{134}} \Big|_{a_5=0}  = \rS_4^{\triple{134}}\,\rV_4^{\triple{134}}\,.
\end{split}
\ee
Gathering  \eqref{p1_p2_f4}  and \eqref{v5_reduction} together and taking into account the symmetry properties of $\mathrm{P}_{1,2}$ \eqref{p1_symm}-\eqref{p2_symm}, one can see that the right-hand side of the reduced pentagon integral \eqref{i5_limit_2} reproduces the right-hand side of the box integral \eqref{i4_result} so that   the desired reduction condition \eqref{i5_limit_1} holds true. 

The other cases $a_l=0$ for $l=1,2,3,4$ can be analyzed in a similar way. The only difference is which of  basis functions $\BF_5^{\triple{ijk}}$ survive at $a_l=0$, but our way of numbering   basis functions immediately indicates  that these are  functions with   upper indices $i,j,k \neq l$. Thus, our expression for the pentagon integral \eqref{i5_result} consistently reduces to the box integral formula \eqref{i4_result}.

\subsection{Hexagon integral}
\label{sec:hex}
Let us finally examine the reconstruction idea for  the 6-point parametric conformal integral. The first bare integral \eqref{in1sums} is given by 
\be
\begin{split}
\label{i61sums}
\cI_6^{(1),{\bm a}} &= N_6^{\bm a} \,
\Gamma \big( 
1-\alpha_{5,6}, \alpha_{5,6}\big) \, v_1^{a_1} v_2^{a_2} v_3^{a_3}  \sum_{\{ k_l,p_{ij}\} =0}^{\infty} 
\frac{u_1^{k_1}}{k_1!}\,
\frac{u_2^{k_2}}{k_2!}\,
\frac{u_3^{k_3}}{k_3!}\,
\frac{w_{12}^{p_{12}}}{p_{12}!}  \,
\frac{w_{13}^{p_{13}}}{p_{13}!}  \,
\frac{w_{23}^{p_{23}}}{p_{23}!}  
\\
&\times
\Gamma \Bigg[ 
\begin{array}{l l}
a_4 +|\bm k|, a'_5+|\bm k|+|\bm p|, a'_6+|\bm k|+|\bm p|, \cB_1, \cB_2, \cB_3 \\
\qquad   1-\alpha_{5,6}+|\bm k|+|\bm p|, \qquad \cC
\end{array}
\Bigg]\;
\\
&\times
F_D^{(3)} \Bigg[
\begin{array}{l l}
a'_5+|\bm k|+|\bm p|\,,  \cB_1\,, \cB_2\,, \cB_3 \\
\qquad \qquad \cC
\end{array}\bigg|\, 1 - v_1\,, 1 - v_2\,, 1-v_3
\Bigg],
\end{split}
\ee
where $|\bm k| = k_1+k_2+k_3$, $|\bm p| = p_{12}+p_{13}+p_{23}$, and the parameters are encoded as
\be
\ba{c}
\dps
\cB_1= a_1+k_1+p_{12}+p_{13}\,, \quad \cB_2 = a_2+k_2 + p_{12}+ p_{23}\,, \quad \cB_3 = a_3+k_3 + p_{13} + p_{23}\,, 
\\ 
\dps
\cC  = a_1+a_2+a_3+a_4 + 2 (|\bm k| +|\bm p|)\,.
\ea
\ee
The nine cross-ratios in \eqref{i61sums} are 
\be
\label{cross_6}
\ba{ccc}
\dps
u_1 = \dps \frac{X_{14} X_{56}}{X_{15} X_{46}} \,, & \dps u_2 = \frac{X_{24} X_{56}}{X_{25} X_{46}}\,, & \dps u_3 = \frac{X_{34} X_{56}}{X_{35} X_{46}}\,, 
\vspace{2mm}
\\
\dps
w_{12} = \frac{X_{12} X_{45} X_{56} }{X_{15} X_{25} X_{46}}\,, & \dps \quad w_{13} = \frac{X_{13} X_{45} X_{56} }{X_{15} X_{35} X_{46}}\,, & \dps \quad w_{23} = \frac{X_{23} X_{45} X_{56} }{X_{25} X_{35} X_{46}}\,, 
\vspace{2mm}
\\
\dps
v_1 \dps = \frac{X_{16} X_{45}}{X_{15} X_{46}} \,, & \dps v_2 = \frac{X_{26} X_{45}}{X_{25} X_{46}} \,,& \dps v_3 = \frac{X_{36} X_{45}}{X_{35} X_{46}} \,.
\ea
\ee

In order to reconstruct the full hexagon conformal integral  we represent  the first bare integral \eqref{i61sums} near $\bm \eta = \bm 0$  by means of the analytic continuation formula \eqref{in1analyt}:
\be 
\label{i61_analyt}
\cI_6^{(1),{\bm a}}(\bm \eta) = \mathrm{U}_6^{(0),{\bm a}}(\bm \eta)+ \mathrm{U}_6^{(1),{\bm a}}(\bm \eta) +  \mathrm{U}_6^{(2),{\bm a}}(\bm \eta)+ \mathrm{U}_6^{(3),{\bm a}}(\bm \eta)\,,
\ee
where each term can be read off from \eqref{varphi_n^0}-\eqref{varphi_n^q}.   According to the general formula \eqref{varphi_cyclic_relation}, two of  four terms  here  are related as
\be
\label{varphi6_01}
C_6 \circ \Big( L_6^{\bm a}(\bm x) \, \mathrm{U}_6^{(0),{\bm a}}(\bm \eta)\Big) = L_6^{\bm a}(\bm x)\, \mathrm{U}_6^{(1),{\bm a}}(\bm \eta) \,,
\ee
where $C_6 = (123456)\in \cS_6$.  Again, $\mathrm{U}_6^{(0),{\bm a}}$, $\mathrm{U}_6^{(2),{\bm a}}$, $\mathrm{U}_6^{(3),{\bm a}}$ can be chosen as master functions,   while the corresponding  basis functions $\BF_6^{\triple{ijk}}$ are generated from them by cycle permutations. Since the  respective cyclic group $\mathbb{Z}_6$ is of order six, then the three master functions generate a set of 18 basis functions.

At this point, we could suggest  that the 6-point conformal integral is to be constructed as the $\mathbb{Z}_6$-invariant sum of 18 basis functions. If so, one can consider a reduction to the pentagon integral which in its turn has  10 terms \eqref{i5_result}. However, it can be shown that  18 basis functions of the hexagon integral  will suffice to reproduce only 9 basis functions of the pentagon integral. This happens if any of the hexagon  propagator powers is set to zero, $a_l=0$ for  some $l=1,...,6$. 

Assuming that we are correct in our conjecture of reconstruction this means that there should be additional master functions which are not directly  seen from the analytic continuation formula for the first bare integral \eqref{i61_analyt}. Indeed, one can show that available master  and  basis functions turn out to  possess a number of structural properties which allow one to introduce  and systematically describe a complete set of functions needed for reconstruction \cite{Alkalaev:2025zhg}. This set includes both already known functions and those that are missing, as in the hexagon case discussed above.

\section{Reconstruction conjecture}
\label{sec:recon}

Summing up our discussion in the previous sections we can now describe the reconstruction procedure in more detail. 

\begin{itemize}
\item {\it Bipartite representation.}  By formal identical transformations the conformal integral $I_n^{{\bm a}}({\bm x})$ is split into two integral parts that defines  the bipartite Mellin-Barnes representation:
\be
\label{1}
I_n^{{\bm a}}({\bm x}) = L_n^{\bm a}(\bm x)\, \cI_n^{(1),{\bm a}}({\bm \eta}) + L_n^{\bm a}(\bm x)\,  \cI_n^{(2),{\bm a}}({\bm \eta})\,.
\ee
The first term here is found in the form of multivariate generalized  hypergeometric series \eqref{in1sums}, while the second term is represented through the $(n-3)(n-2)/2$--folded Mellin-Barnes integral \eqref{in2mb} with a complex structure of poles, which currently is not known in analytic form. It is the reconstruction procedure  that suggests avoiding any further integration because  the second bare integral can be recovered from the first one  by invoking  the permutation invariance of the full conformal integral.

\item {\it Coordinate domains.} The function $I_n^{{\bm a}}({\bm x})$ is defined on a domain $D_n \subset (\RR^D \times)^n$, which can be found  by  first calculating  the conformal integral explicitly in a particular coordinate domain  and then using  analytic continuation formulas.\footnote{This can be contrasted  with other possible calculation schemes  such as the geometric approach where the conformal integral  is given by volume of a simplex in a space of constant curvature which sign depends on a particular kinematic regime, i.e. on a particular coordinate domain, see e.g. \cite{Bourjaily:2019exo}.}

The two terms in \eqref{1} are supported  on their own domains which are generally wider than $D_n$. The domain of the first term $D_n^{(1)}$ is such that $D_n \subset D_n^{(1)}$, it is also hard to identify explicitly. We denote the respective domain  in the space of cross-ratios $\cH_{n}$ as $\widehat{D}^{(1)}_n \subset \cH_{n}$, i.e. $\widehat{D}^{(1)}_n = {\bm \eta} (D_n^{(1)})$.

\begin{itemize}
\item There is a smaller domain $\widehat{A}_n$ which is the convergence region of $\cI_n^{(1),{\bm a}}$ calculated as a multivariate power series near a particular point:
\be
\label{2}
\widehat{D}^{(1)}_n \supset \widehat{A}_n \;\ni\; ({\bm u}=0, {\bm w} = 0, {\bm v}= 1)  \,.
\ee

\item Using the known  analytic continuation formulas one can define  $\cI_n^{(1),{\bm a}}$ on  a different domain $\widehat{B}_n$ containing the origin of coordinates in $\cH_n$:  
\be
\label{3}
\widehat{D}^{(1)}_n  \supset \widehat{B}_n \;\ni\; ({\bm u}=0, {\bm w}= 0, {\bm v}= 0) \,.
\ee

\item In fact, an exact shape of either domain remains unclear. Instead, we operate  with small enough neighbourhoods of concrete points in $\cH_n$ like $0,1,\infty$.
\end{itemize}

\item {\it Symmetric group.}  By construction, the conformal integral is invariant under action of the symmetric group
\be
\label{4}
\forall \pi \in \cS_n:\quad \mathrm{R}(\pi)\circ   I_n^{{\bm a}}({\bm x}) = I_n^{{\bm a}}({\bm x})\,.
\ee
There are maps of the (co)domains induced by elements of the symmetric group $\cS_n$ acting on ${\bm x} = (x_1, ..., x_n)$. In the space of cross-ratios $\cH_n$ the symmetric group acts by  transformations 
\be
\label{5}
\forall \pi \in \cS_n:\quad  {\bm \eta} \to {\bm \eta}' =  \mathrm{R}(\pi) \circ{\bm \eta}  =   H_\pi({\bm \eta})\,,
\ee
where $H_\pi$ are homogeneous rational functions. One can single out various subgroups of $\cS_n$ which act in $\cH_n$ in some special way. 

\begin{itemize}

\item E.g., one can consider a stabilizer (kinematic) group, 
\be
\label{6}
\cS_n^{{\rm kin}}:= \{ \pi \in \cS_n: \; \mathrm{R}(\pi) \circ {\bm \eta}  = {\bm \eta},\, \forall {\bm \eta} \in \cH_n\} \subset \cS_n\,.
\ee
It is clear that the stabilizer leaves  any domain in $\cH_n$ invariant. However, one can show that  the stabilizers are all trivial for $n>4$ \cite{Kravchuk:2016qvl}.

\item Other subgroups change chosen  domains. Given a domain $\widehat{X}_n \subset \cH_n$ one can consider a subgroup  
\be
\label{7}
\cG_n \subset \cS_n:\quad \widehat{X}_n^\prime = \mathrm{R}(\cG_n) \circ \widehat{X}_n
\quad \text{such that} \quad 
\widehat{X}_n \cap \widehat{X}_n^\prime \neq \varnothing   \,.
\ee     
E.g., there is a cyclic subgroup generated by  $C_n = (123...n)\in \cS_n$, i.e. 
\be
\label{8}
\mathbb{Z}_{n} = \{e, C_n, (C_n)^2, ..., (C_n)^{n-1}\} \subset \cS_n\,.
\ee

\item One may consider extended kinematic groups $\widehat \cS_n^{{\rm kin}}$ which are nontrivial for $n>4$ and coincide with $\cS_4^{{\rm kin}}$ when $n=4$ (see appendix \bref{app:kin}). However, at our current level of understanding, $\widehat \cS_n^{{\rm kin}}$ are derived objects, i.e. these subgroups can be defined only in terms of the cyclic group and its action on basis functions that is not useful in practice.  
\end{itemize}

\item {\it Basis functions.}  One can choose a set of {\it basis} functions $\BF_n^{\triple{ijk}}$ supported on a common domain $\widehat{X}_n \subset \widehat{D}_n$.  The considered cases $n=4,5,6$ allow us to emphasize that the basis functions share a number of  common features:
\begin{itemize}
\item $\Gamma$-function prefactor $\rS_n^{\triple{ijk}}(\bm a)$\;;

\item Leg-factor $\rV_n^{\triple{ijk}}(\bm a|\bm x)$\;;

\item Hypergeometric-type power series in $n(n-3)/2$ cross-ratios.

\end{itemize}

Moreover, each of these elements  and, therefore, a given basis function is  determined by choosing 3 pairs $(x_i,a_i)$ out of $n$ such pairs.

\item {\it Master functions.}  The conformal integral $I_n^{{\bm a}}({\bm x})$ must be invariant under the symmetric group $\cS_n$. However, the basis functions $\BF_n^{\triple{ijk}}(\bm a|\bm x)$ are supported on a particular coordinate domain that  implies that the full symmetric group is broken down to a particular {\it generating} subgroup of the type \eqref{7} which acts on the basis functions in a specific way: among all basis functions one can single out a number of {\it master} functions which generate all other basis functions by acting with elements of the generating group. By construction, the generating group reshuffles a set of basis functions. It is important to stress that a set of basis functions depends on the choice of a coordinate domain that in its turn defines a suitable generating group. 

\item {\it Consistency checks.} In order to check the resulting expressions  for the conformal integrals obtained by the reconstruction method  one can compare them with those known in the  literature. There are two  possible checks: (1) reducing a number of points, $n \to n-1$; (2) going to the non-parametric case, $a_i\to 1$.

\end{itemize} 

\noindent If one chooses a domain as $\widehat{X}_n  = \widehat{B}_n$ and a generating group as $\mathbb{Z}_n$,  then  the conformal integral can explicitly be evaluated by means of the following

\paragraph{Reconstruction conjecture.} {\it The $n$-point conformal integral \eqref{indef} is the $\mathbb{Z}_n$-invariant sum of basis functions
\be
\label{conj_fin}
I_n^{{\bm a}}({\bm x})\, \overset{\mathrm{nr}}{=}\; \sum_{m=0}^{n-1} (\cycle_n)^m  \circ \sum_{\triple{ijk} \in {\rm T}_n} \BF_n^{\triple{ijk}}(\bm a|\bm x)\,,
\ee
where ${\rm T}_n$ is a set of index  triples $\triple{ijk}$ which are not related to each other by cyclic permutations, $1\leq i<j<k \leq n$. A cardinal number of ${\rm T}_n$ equals the number of master functions,
\be
\label{cord}
\left| {\rm T}_n \right| = \begin{cases}
\dps \frac{(n-2)(n-1)}{3!}\,, \quad \text{if} \quad \frac{n}{3} \notin \mathbb{N}\,, 
\vspace{2mm}
\\
\dps \frac{(n-2)(n-1) + 4}{3!} \,, \quad \text{if} \quad \frac{n}{3} \in \mathbb{N}\,.
\end{cases}
\ee 
A total number of basis functions is equal to the binomial coefficient 
\be
\label{num_basis}
\binom{n}{3} = \frac{n(n-1)(n-2)}{3!}\,,
\ee
which determines a number of all possible index triples. The symbol $\overset{\mathrm{nr}}{=}$ in \eqref{conj_fin} means that among all basis functions produced by acting with cycles on master functions one keeps only non-repeating ones which number equals \eqref{num_basis}. 
}

As drafted, this conjecture  shifts the focus to finding master functions as well as describing their properties. A few comments are in order.

\begin{itemize}
\item Some of  master functions are contained directly in the calculated part $L_n^{\bm a}(\bm x)\, \cI_n^{(1),{\bm a}}$.  Since among $n-2$ candidate master functions delivered by the analytic continuation formula  \eqref{in1analyt} the first two functions are cyclically dependent (see \eqref{varphi_cyclic_relation}), there remain $n-3$ master functions which we explicitly know from evaluating the first bare integral. However, a number of master functions is conjectured to grow quadratically \eqref{cord} and $n=6$  is the first time when the $n-3$ functions turn out to be  insufficient for the full reconstruction. 

\item The  number of master and basis functions \eqref{cord} and \eqref{num_basis} in the considered cases $n=4,5,6$ equals  $1$, $2$, $4$ and $4$, $10$, $20$, respectively. In the hexagon case we see that in addition  to 18 basis functions coming  from the first bare integral there are two more basis functions produced from the additional master function\footnote{This means that the respective orbit of the cyclic group $\mathbb{Z}_6$ has length 2 instead of 6 as in the case of other hexagon master functions. This  orbit shortening happens when $n/3 \in \mathbb{N}$.} and now their total number is sufficient to reproduce the pentagon integral when checking  the $n \to n-1$ reduction.

\item  We claim that unknown  part of master functions can be derived  from already known  basis functions. The point is that the proposed  parameterization by index triples allows one to reveal a series of  remarkable properties of master/basis functions which can be effectively described by means of some diagrammatics. This will enable  us eventually  to build a complete set of functions explicitly, thereby reconstructing  the full conformal integral in a given coordinate domain \cite{Alkalaev:2025zhg}.   

\end{itemize}

\section{Conclusion}
\label{sec:conclusion}

In this paper, we have proposed a new method of calculating $n$-point parametric one-loop conformal integrals and elaborated two examples of the box and pentagon (non)-parametric  integrals. Essentially, there are three main points: (1) a bipartite Mellin-Barnes representation which allows one to evaluate an additive  part of the conformal integral explicitly; (2)  asymptotic expansions near particular points and analytic continuation on other domains; (3) permutation invariance can be used to recover the remaining unknown part.

This reconstruction procedure operates with a set of master functions which generate a wider set of basis functions by means of permutations from the  cyclic group. The resulting expression for the conformal integral supported on a particular coordinate domain is given  by summing all  basis functions. For the box and pentagon integrals we have shown  that the basis functions are directly defined from the explicitly found  part of the bipartite representation. For the hexagon and higher-point integrals a complete set of basis functions will be described  in  our  forthcoming paper \cite{Alkalaev:2025zhg} in which will finalize the reconstruction method. 

Note that to some extent our approach parallels the Yangian bootstrap proposed in \cite{Loebbert:2019vcj}. Indeed, there are two ingredients in the Yangian bootstrap  which are the permutation symmetry and $n\to n-1$ reduction  that allows one to partially fix the solution space of the Yangian invariance equations satisfied by the $n$-point conformal integrals. It also allows one to  consider its analyticity regions where the integral can be represented by one or another multivariate series. In contrast, the reconstruction method captures a particular analyticity domain from the very beginning that, from the Yangian bootstrap perspective, allows one to significantly reduce a number of power series and their constant prefactors which can represent the conformal integral in a given domain.

It would be natural to find an appropriate modification of the reconstruction method   for at least two types of Feynman integrals possessing conformal symmetry. These are  multi-loop higher-point conformal integrals which are currently being considered by other methods, see e.g. \cite{Paulos:2012nu,Basso:2017jwq,Derkachov:2018rot,Duhr:2023bku,Loebbert:2024fsj, Olivucci:2021cfy, Derkachov:2021ufp, Olivucci:2023tnw, Aprile:2023gnh} as well as  massive conformal integrals studied in \cite{Bourjaily:2019exo, Loebbert:2020hxk, Loebbert:2020glj}. The latter have recently been shown to play an essential role when calculating contact Witten diagrams \cite{Rigatos:2022eos}. 

Besides possible generalizations, the already obtained results can be immediately applied in \cft by means of the shadow formalism (see our discussion in the Introduction). Especially, the one-loop conformal integrals are useful in \cft on non-trivial backgrounds, where one can show that the problem of calculating $n$-point thermal conformal blocks boils down to knowing $(2n+2)$-point one-loop conformal integrals explicitly (see \cite{Alkalaev:2024jxh} for 1-point thermal blocks). 

\vspace{3mm} 

\noindent \textbf{Acknowledgements.} We would like to thank Leonid  Bork, Sergey Derkachev, Alexey  Isaev for discussions about related topics, and  Marcus Spradlin for correspondence.
S.M. is also very grateful to Ekaterina Semenova for her support at all stages of this project as well as for her help in drawing diagrams. 

Our work was supported by the Foundation for the Advancement of Theoretical Physics and Mathematics “BASIS”.

\appendix

\section{Notation and conventions}
\label{app:notations}

Let $\bm g = \{g_1,..., g_N \}$ denote a set of elements such as parameters of functions, powers, integration variables, etc. Their various sums will be denoted as
\be 
\label{notation_summations}
|\bm g_{i,j}| = \sum_{l=i}^j g_l\,, \quad |\bm g| = \sum_{l=1}^N g_l\,, \quad |\bm g_q| = |\bm g_{q,N}| - |\bm g_{1,q-1}|\,.
\ee
In the case of propagator powers  $\bm a  = (a_1,..., a_n)$ we also use
\be
\label{notations_params}
|\bm a_{i,j}| = \begin{cases}
				\dps
				\sum_{l=i}^j a_l \,, \quad i<j\,, \vspace{2mm }\\
				\dps
				\sum_{l=i}^n a_l + \sum_{l=1}^j a_l\,, \quad i>j\,,
				\end{cases} 
\quad
a'_i = \hd - a_i\,,  \quad \alpha_{i,j} = a_i + a_j -\hd \equiv a_i -a_j'\,,
\ee
where $D$ is the dimension of Euclidean  space  $\mathbb{R}^D$.

The products of $\Gamma$-functions are denoted as 
\be
\label{gammasprod}
\Gamma \Bigg[ 
\begin{array}{l l}
a_1, ..., a_M \\
b_1, ..., b_K
\end{array}
\Bigg] = \frac{\Gamma(a_1,..., a_M)}{\Gamma(b_1,..., b_K)} \,, \qquad \Gamma(a_1,..., a_M) = \prod_{i=1}^M \Gamma(a_i)\,.
\ee
In the Mellin-Barnes integrals it is convenient to use  the following modified measure:
\be 
\label{mb_measure}
\widehat{{\rm d} t} = \frac{{\rm d} t}{2 \pi i} \, \Gamma(-t)\,.
\ee

\section{Generalized hypergeometric  functions}
\label{app:functions}
In this appendix we discuss various  generalized hypergeometric functions focusing on their integral representations, identities as well as analytic continuation formulas.  We closely follow the review   \cite{Bezrodnykh:2018review}. See also \cite{Bateman:100233, Olsson:1964, Dubovyk:2022obc, Bezrodnykh:2018acl}. 
 
\subsection{General definitions}
Following the Horn's classification scheme, a $N$-point  multivariate  power series 
\be
\label{generalseries}
\sum_{m_1,\ldots, m_N=-\infty}^{+\infty} A(m_1,..., m_N)\, \xi_1^{m_1} ...\, \xi_N^{m_N} \equiv \sum_{{\bm m} \in \mathbb{Z}^N} A({\bm m})\, {\bm \xi}^{\bm m}
\ee
is a generalized hypergeometric function  if
\be
\label{horncond}
\frac{A({\bm m}+{\bm e}_j)}{A({\bm m})} = \frac{P_j ({\bm m})}{Q_j ({\bm m})}\,, \qquad   \forall j=1,...\,, n\,, 
\ee
where $P_j({\bm m})$ and $Q_j({\bm m})$ are some polynomials in ${\bm m}=\{m_1,..., m_N \}$ and ${\bm e}_j = \{0,...,1,...,0 \}$ is a vector with all components equal to zero except $j$-th one. In other words, a given generalized hypergeometric function \eqref{generalseries} have expansion coefficients such that their ratios are rational functions of parameters.    

The relation \eqref{horncond} treated as an equation for  $A(m_1,..., m_N)$ admits many solutions. In particular, when 
\be
\label{fdpolinoms}
P_j ({\bm m}) = (a + |{\bm m}|)(b_j + m_j)\,, \qquad Q_j ({\bm m}) = (c + |{\bm m}|)(1 + m_j)\,,
\ee
where ${\bm m}, {\bm b}, a, c$ are constant complex-valued parameters (see \eqref{notation_summations} for notation), the solution to \eqref{horncond} is the $N$-point Lauricella function $D$:
\be
\begin{split}
\label{fd}
F_{D}^{(N)} \Bigg[ 
\begin{array}{l l}
a\,, {\bm b} \\
\,\, c
\end{array}\bigg|
{\bm \xi}
\Bigg] &= \sum_{|{\bm m}|=0}^{\infty} \frac{(a)_{|{\bm m}|} ({\bm b})_{\bm m}}{(c)_{|{\bm m}|}} 
\frac{{\bm \xi}^{\bm m}}{{\bm m}!} \,,
\end{split}
\ee
which is convergent  in the unit polydisk
\be
\label{polydisk}
\mathbb{U}^{N} = \left\lbrace {\bm \xi} \in \mathbb{C}^N: |\xi_j| < 1\,,\, j=1,...,N \right\rbrace \,.
\ee
The (rising) factorials in  \eqref{fd} are defined as 
\be
(a)_m = \frac{\Gamma(a+m)}{\Gamma(a)}\,, \qquad ({\bm b})_{\bm m} = (b_1)_{m_1}\,...\, (b_N)_{m_N}\,, \qquad {\bm m}! = m_1!\,...\, m_N!\,.
\ee
At $N=1$ and $N=2$ the Lauricella function $D$ \eqref{fd} becomes the Gauss hypergeometric function ${}_2 F_1$ and the first Appell function $F_1$, respectively:
\be
\label{2f1}
{}_2 F_1 \Bigg[ 
\begin{array}{l l}
a\,, b \\
\,\, c
\end{array}\bigg|
\xi
\Bigg] = \sum_{m=0}^{\infty} \frac{(a)_m (b)_m }{(c)_m} \frac{\xi^m}{m!} \,,
\ee
\be
\label{f1}
F_1 \Bigg[ 
\begin{array}{l l}
a, b_1, b_2\\
\quad c
\end{array}\bigg|
\xi_1, \xi_2
\Bigg] = \sum_{m_1,m_2 = 0}^{\infty} \frac{(a)_{m_1+m_2} (b_1)_{m_1}  (b_2)_{m_2}}{(c)_{m_1+m_2}  } \frac{\xi_1^{m_1} }{m_1!} \frac{\xi_2^{m_2} }{m_2!} \,.
\ee

The Lauricella function $F_D^{(N)}$ belongs to the family of generalized hypergeometric functions $G^{(N,j)}$ labelled  by $j=1,...,N+1$ (see \eqref{notation_summations} for notation):
\be
\label{gnj}
G^{(N,j)} \Bigg[ 
\begin{array}{l l}
a\,, {\bm b} \\
\,\, c
\end{array}\bigg|
{\bm \xi}
\Bigg] = \sum_{|{\bm m}|=0}^{\infty} \frac{(a)_{|{\bm m_j}|} ({\bm b})_{\bm m}}{(c)_{|{\bm m_j}|}} 
\frac{{\bm \xi}^{\bm m}}{{\bm m}!} \,,
\ee
which converge in the unit polydisk \eqref{polydisk}. The subscripts of the Pochhammer symbols in \eqref{gnj} can be negative, in which case the Pochhammer symbol reads as
\be
\label{pochhnegative}
(a)_{-m} = \frac{(-1)^m}{(1-a)_m}\,, \quad m\in \mathbb{Z}^+\,.
\ee 
The Lauricella function $F_D^{(N)}$ follows from  \eqref{gnj} at $j=1$ and $j=N+1$:
\be
\label{gnjfd}
F_D^{(N)} \Bigg[ 
\begin{array}{l l}
a\,, {\bm b} \\
\,\, c
\end{array}\bigg|
{\bm \xi}
\Bigg] = G^{(N,1)} \Bigg[ 
\begin{array}{l l}
a\,, {\bm b} \\
\,\, c
\end{array}\bigg|
{\bm \xi}
\Bigg]\,, 
\qquad 
F_D^{(N)} \Bigg[ 
\begin{array}{l l}
1-c\,, {\bm b} \\
\,\, 1-a
\end{array}\bigg|
{\bm \xi}
\Bigg]=G^{(N,N+1)} \Bigg[ 
\begin{array}{l l}
a\,, {\bm b} \\
\,\, c
\end{array}\bigg|
{\bm \xi}
\Bigg]\,.
\ee
The Horn function $G_{2}$ follows from \eqref{gnj} at $j=2$ and $N=2$:
\be
\label{g2}
\ba{l}
\dps
G^{(2,2)} \Bigg[
\begin{array}{l l}
a, b_1, b_2 \\
\quad c
\end{array}\bigg| \xi_1,\xi_2
\Bigg]  = G_2\left(b_1,b_2,a,1-c\big| -\xi_1, -\xi_2 \right) 
\vspace{2mm} 
\\
\dps
\hspace{30mm}= \sum_{m_1,m_2=0}^{\infty}  (b_1)_{m_1} (b_2)_{m_2} (a)_{m_2-m_1} (1-c)_{m_1-m_2}\frac{(-\xi_1)^{m_1}}{m_1!}  \frac{(-\xi_2)^{m_2}}{m_2!} \,.
\ea
\ee

Finally, note that given a $N$-point generalized hypergeometric function one can construct a $(N+m)$-point hypergeometric function through the so-called splitting identities. E.g., the fourth Appell function $F_4$ defined by the double power series
\be
\label{f4}
F_4 \Bigg[
\begin{array}{l l}
a_1,a_2  \\
c_1,c_2
\end{array}\bigg| 
\xi_1, \xi_2
\Bigg] 
= 
\sum_{m_1,m_2 = 0}^{\infty} \frac{(a_1)_{m_1+m_2} (a_2)_{m_2+m_1}  }{(c_1)_{m_1} (c_2)_{m_2} } \frac{\xi_1^{m_1} }{m_1!} \frac{\xi_2^{m_2} }{m_2!} \,,
\ee
and converging when $\sqrt{\xi_1}+\sqrt{\xi_2} < 1$ can be represented  in terms of the Gauss hypergeometric function \eqref{2f1} by means of  the following splitting identity 
\be
\label{f4split}
F_4 \Bigg[
\begin{array}{l l}
a_1,a_2 \\
c_1,c_2
\end{array}\bigg| 
\xi_1,\xi_2
\Bigg] = \sum_{m_1=0}^{\infty} \frac{\xi_1^{m_1}}{m_1!} \frac{(a)_{m_1} \, (a_2)_{m_1} }{(c)_{m_1}}\; {}_2 F_1\Bigg[
\begin{array}{l l}
a_1+m_1, a_2+m_2  \\
\qquad c_2
\end{array}\bigg| 
\xi_2
\Bigg].
\ee

\subsection{Integral representations}

Generalized hypergeometric functions can also be defined through the Euler-type integrals. E.g., for $\operatorname{Re}b_i >0\,, \operatorname{Re}(c-|{\bm b}|) >0$ the Lauricella function  $F_D^{(N)}$ \eqref{fd} can be represented as
\be
\label{fdint}
F_{D}^{(N)} \Bigg[ 
\begin{array}{l l}
a\,, {\bm b} \\
\,\, c
\end{array}\bigg|
{\bm \xi}
\Bigg] 
= \Gamma\Bigg[
\begin{array}{l l}
\quad c\\
c-|{\bm b}|, {\bm b}
\end{array}
\Bigg] \,  
\int_{\Omega} {\rm d}^{N}\sigma  \,\frac{{\bm \sigma}^{\bm b- \bm 1} (1-|{\bm \sigma}|)^{c-|{\bm b}|-1} }{(1-{\bm \sigma} \cdot {\bm \xi})^a }\;,
\ee
where ${\bm \sigma} \cdot {\bm \xi} = \sigma_1 \xi_1 + ... + \sigma_N \xi_N$, and the integration domain is the standard orthogonal simplex $\Omega = \{{\bm \sigma} \in \mathbb{R}^N: \sigma_j \geq 0\,, \, j=1,...,N\,;\, |{\bm \sigma}|\leq 1 \}$ (see  \eqref{gammasprod} for notation).

Euler-type integrals are useful in finding  transformation formulas for generalized  hypergeometric functions. This can be illustrated by another integral representation of the Lauricella function $F_D^{(N)}$:
\be
F_{D}^{(N)} \Bigg[ 
\begin{array}{l l}
a\,, {\bm b} \\
\,\, c
\end{array}\bigg|
{\bm \xi}
\Bigg] = \Gamma
\Bigg[
\begin{array}{l l}
\quad c\\
c-a, a
\end{array}
\Bigg] \,  \int_0^{\infty}{\rm d}t \, t^{a-1} (1-t)^{c-a-1} \prod_{j=1}^N (1-t\, \xi_j)^{-b_j} \,,
\ee
valid when $\operatorname{Re}a >0$, $\operatorname{Re}(c-a) >0$, and $|\operatorname{arg}(1-\xi_j)| < \pi$, $\forall j=1,...,N$. Changing the integration variable as $t \to 1-t$ one can establish the following transformation 
\be
\begin{split}
\label{laurpfaff}
F_{D}^{(N)} \Bigg[ 
\begin{array}{l l}
a\,, {\bm b} \\
\,\, c
\end{array}\bigg|
{\bm \xi}
\Bigg] =  ({\bm 1- \bm \xi})^{-{\bm b}} \, F_{D}^{(N)} \Bigg[ 
\begin{array}{l l}
c-a\,, {\bm b} \\
\quad c
\end{array}\bigg|
\mathcal{P}({\bm \xi})
\Bigg]\,, \quad \mathcal{P}({\bm \xi}) = \left\lbrace \frac{\xi_1}{\xi_1 - 1 }\,,..., \frac{\xi_N}{\xi_N-1} \right\rbrace\,.
\end{split}
\ee
At $N=1$ it is the Pfaff transformation of the Gauss hypergeometric function.

Moreover, the integral representations can be  useful in finding analytic continuation formulas. E.g.  the Gauss hypergeometric function ${}_2 F_1$ has the following integral representation
\be
\label{2f1intinfty}
{}_2 F_1 \Bigg[
\begin{array}{l l}
\qquad a,b  \\
1+a+b-c
\end{array}\bigg| 
1-\xi
\Bigg] = \Gamma 
\Bigg[
\begin{array}{l l}
1+a+b-c \\
\qquad a,b
\end{array}
\Bigg] \, \int_0^{\infty} {\rm d}s\, \frac{s^{b-1} (1+s)^{c-b-1} }{(1+s \, \xi)^{a}}\,,
\ee
which extends the original power series definition to a domain around 1. 

\paragraph{Mellin-Barnes integrals.} In order to deal with the problem of analytic continuation of generalized hypergeometric functions, it is useful to use the Mellin-Barnes integral technique. E.g. the Mellin-Barnes representation for  the Gauss hypergeometric function is given by  
\be
\label{2f1mb}
{}_2 F_1 \Bigg[
\begin{array}{l l}
a,b  \\
\,\, c
\end{array}\bigg| 
\xi
\Bigg]  = \Gamma
\Bigg[
\begin{array}{l l}
\,\, c \\
a,b
\end{array}
\Bigg] \, \int_{-i\infty}^{+i\infty}  \frac{{\rm d} s}{2\pi i} \, 
\Gamma
\Bigg[
\begin{array}{l l}
a+s,b+s, -s \\
\qquad c+s
\end{array}
\Bigg] (-\xi)^s \,,
\ee
where the integration contour is chosen such  that the poles of the $\Gamma$-functions $s^{(1)}=-a-k$ and $s^{(2)}=-b-k$ lie to the left of the contour and  the poles $s^{(3)}=k$ lie to the right ($k \in \mathbb{Z}^+$).

The Mellin-Barnes integrals are also useful in representing elementary functions, e.g. the exponential function
\be
\label{mbexponent}
\exp \left(-X\right)  = \frac{1}{2\pi i}\int_{-i \infty}^{+i \infty} {\rm d}s\, \Gamma(-s) X^s \,,
\ee
and the power function
\be
\begin{split}
\label{mbexpansion}
\frac{1}{(\lambda_1 + ...+ \lambda_N)^a} &= \frac{1}{\Gamma(a)}\, \frac{1}{(2 \pi i)^{N-1}}\, \int_{-i \infty}^{+ i \infty} {\rm d}s_1 \cdots \int_{-i \infty}^{+ i \infty} {\rm d}s_{N-1}  \Gamma(-s_{1}) \cdots  \Gamma(-s_{N-1})\\
&\times \Gamma(s_1+... + s_{N-1}+a) \, \lambda_1^{s_1} \cdots \lambda_{N-1}^{s_{N-1}} \, \lambda_n^{-s_1-\ldots -s_{N-1}-a}\,.
\end{split}
\ee
In the last formula the integration contours for $s_i$ are chosen such that poles $s_i^{(1)} = -S_i-k_i$ are to the left, where $S_i=s_1+...+s_N+a-s_i$ and poles $s_i^{(2)}=-k_i$ are to the right, $k_i \in \mathbb{Z}^+\,, \, \forall i=1,...,N-1$.

\subsection{Analytic continuation}
When calculating Mellin-Barnes integrals, e.g. \eqref{2f1mb}, there is an ambiguity in  closing a contour  to the left or right. However, it can be  fixed by specifying domains of  argument $\xi$ and parameters $a,b,c$.  More precisely, when closing a contour to the right, the integral \eqref{2f1mb} can be calculated as a sum of residues at the poles $s^{(3)}=k\in \mathbb{Z}^+$, yielding the Gauss hypergeometric series \eqref{2f1} which  converges for  $|\xi|<1$. On the other hand, when closing  a contour to the left, the same integral is a sum of residues at the poles $s^{(1)}=-k-a$ and $s^{(2)}=-k-b$ with  $k\in \mathbb{Z}^+$.  In this way, one obtains the analytic continuation formula for the Gauss hypergeometric series \eqref{2f1} in the domain $|\xi| > 1$:
\be
\begin{split}
{}_2 F_1 \Bigg[ 
\begin{array}{l l}
a\,, b \\
\,\, c
\end{array}\bigg|
\xi
\Bigg] &= \Gamma
\Bigg[
\begin{array}{l l}
c, b-a \\
b, c-a
\end{array}
\Bigg]\, (-\xi)^{-a}  {}_2 F_1 \Bigg[ 
\begin{array}{l l}
a, 1+a-c \\
\,\, 1+a-b
\end{array}\bigg|
\frac{1}{\xi}
\Bigg] \\
&+ 
\Gamma
\Bigg[
\begin{array}{l l}
c, a-b \\
a, c-b
\end{array}
\Bigg] (-\xi)^{-b} {}_2 F_1 \Bigg[ 
\begin{array}{l l}
b, 1+b-c \\
\,\, 1+b-a
\end{array}\bigg|
\frac{1}{\xi}
\Bigg].
\end{split}
\ee
Another analytic continuation formula can be derived by means of the first Barnes lemma:
\be
\intbrom \frac{{\rm d}t}{2 \pi i} \Gamma(a+t, b+t, s-t, c-a-b-t) = \Gamma
\Bigg[
\begin{array}{l l}
a+s,b+s,c-b,c-a \\
\qquad c+s
\end{array}
\Bigg].
\ee
Substituting this relation into \eqref{2f1mb} one obtains the following integral representation 
\be
\label{2f1mb_2}
{}_2 F_1 \Bigg[
\begin{array}{l l}
a,b  \\
\,\, c
\end{array}\bigg| 
\xi
\Bigg]  = \Gamma
\Bigg[
\begin{array}{l l}
\qquad c \\
a,b,c-a,c-b
\end{array}
\Bigg] \, \int_{-i\infty}^{+i\infty}  \frac{{\rm d} t}{2\pi i} \, \Gamma(a+t,b+t,c-a-b-t,-t)
(1-\xi)^t \,,
\ee
where we also used \eqref{mbexpansion} to evaluate the integral over $s$. Computing the integral as a sum over residues at the  poles $t^{(1)} = k \in \mathbb{Z}^+$ and $t^{(2)} = c-a-b+k$, we find  the following analytic continuation formula for the Gauss hypergeometric function 
\be
\begin{split}
\label{2f1analyt}
{}_2 F_1 \Bigg[
\begin{array}{l l}
a,b  \\
\,\, c
\end{array}\bigg| 
\xi
\Bigg] &= \Gamma
\Bigg[
\begin{array}{l l}
c, c-a-b\\
c-a, c-b
\end{array}
\Bigg]  \, {}_2 F_1 \Bigg[
\begin{array}{l l}
a,b  \\
1+a+b-c
\end{array}\bigg| 
1-\xi
\Bigg]
\\
&+  \Gamma
\Bigg[
\begin{array}{l l}
c, a+b-c\\
a,b
\end{array}
\Bigg]  \, (1-\xi)^{c-a-b} \, {}_2 F_1 \Bigg[
\begin{array}{l l}
c-a,c-b  \\
1+c-a-b
\end{array}\bigg| 
1-\xi
\Bigg]\,.
\end{split}
\ee
By extending  this procedure to generalized hypergeometric functions we also find the analytic continuation formula for the Lauricella function $F_D^{(N)}$ \eqref{fd}:\footnote{For $N=2$, the formula \eqref{fdanalyt} which was first derived in \cite{Olsson:1964}, provides an analytic continuation of the first Appell function $F_1$. }
\be
\label{fdanalyt}
F_{D}^{(N)} \Bigg[ 
\begin{array}{l l}
a\,, {\bm b} \\
\,\, c
\end{array}\bigg|
{\bm \xi}
\Bigg] = \sum_{q=0}^{N} A_q \, \mathrm{D}_q
\Bigg[ 
\begin{array}{l l}
a\,, {\bm b} \\
\,\, c
\end{array}\bigg|
{\bm \xi}
\Bigg]\,,
\ee
with a new domain of convergence    
\be
\mathbb{K}^N =  \left\lbrace {\bm \xi} \in \mathbb{C}^N: 0<|1-\xi_1| < ...< |1-\xi_N|<1\,;\, |{\rm arg}(1-\xi_j)|<\pi\,, j=1,...,N \right\rbrace \,,
\ee
cf. \eqref{polydisk}.  The functions $\mathrm{D}_q$ are given by 
\be
\label{fdanalytfunc}
\begin{split}
\mathrm{D}_0
\Bigg[ 
\begin{array}{l l}
a\,, {\bm b} \\
\,\, c
\end{array}\bigg|
{\bm \xi}
\Bigg] &= F_D^{(N)} \Bigg[ 
\begin{array}{l l}
a\,, {\bm b} \\
1+a+|\bm b|-c
\end{array}\bigg|
\bm 1 - {\bm \xi}
\Bigg]\,, \\
\mathrm{D}_{q}
\Bigg[ 
\begin{array}{l l}
a\,, {\bm b} \\
\,\, c
\end{array}\bigg|
{\bm \xi}
\Bigg] &= (1-\xi_q)^{c-a-|\bm b_{1,q}|} \Big(\prod_{l=q+1}^{N} (1-\xi_l)^{-b_l}\Big) G^{(N,q)} \Bigg[ 
\begin{array}{l l}
c-a-|\bm b_{1,q-1}|\,, \widetilde{\bm b}_q \\
1+c-a-|\bm b_{1,q}|
\end{array}\bigg|
\bm \Xi^{(q)}
\Bigg]\,,
\end{split}
\ee
where $q=1,...,N$ and  $G^{(N,q)}$ is defined in \eqref{gnj}. The parameters and arguments in \eqref{fdanalytfunc} are packed into the following combinations:
\be
\begin{split}
\widetilde{\bm b}_q &= \left\lbrace b_1\,, ..., b_{q-1}\,, c-|\bm b|\,, b_{q+1}\,,..., b_N \right\rbrace\,, \\
\bm \Xi^{(q)} &= \left\lbrace \frac{1-\xi_1}{1-\xi_q}\,,..., \frac{1-\xi_{q-1}}{1-\xi_q}\,, 1-\xi_q \,, \frac{1-\xi_q}{1-\xi_{q+1}}\,, ..., \frac{1-\xi_q}{1-\xi_N} \right\rbrace.
\end{split}
\ee
The coefficients in \eqref{fdanalyt} are given by
\be
\label{fdanalytcoef}
\begin{split}
A_0 &= \Gamma
\Bigg[
\begin{array}{l l}
c,c-a-|\bm b|\\
c-|\bm b|, c-a
\end{array}
\Bigg] , \qquad 
A_{q} = \Gamma 
\Bigg[
\begin{array}{l l}
c, c-|\bm b_{1,q-1}|-a, a+|\bm b_{1,q}|-c \\
a, b_q, c-a
\end{array}
\Bigg].
\end{split}
\ee
It is worth noting that the right-hand side of \eqref{fdanalyt} contains two Lauricella functions. The first one is given by $\mathrm{D}_{0}$, while the term $\mathrm{D}_{1}$ is proportional to the Lauricella function  by virtue of \eqref{gnjfd}.

\section{One more Mellin-Barnes representation of the conformal integral}
\label{app:MB_more}
In this section,  we describe  an alternative way of representing the conformal integral in terms of Mellin-Barnes integrals. To this end, one steps back to \eqref{inbare2f1} and represents the Gauss hypergeometric function ${}_2 F_1$ therein by means of the analytic continuation formula \eqref{2f1mb_2}:
\be
{}_2 F_1 \Bigg[
\begin{array}{l l}
a'_{n-1} ,  a'_n \\
\; \quad \hd
\end{array}\bigg| 
1-\xi({\bm \sigma})
\Bigg] = \intbrom \widehat{{\rm d} t} \,
\Gamma 
\Bigg[
\begin{array}{l l}
\hd,  a'_{n-1}+t, a'_{n}+t, \alpha_{n-1,n}-t \\
a_{n-1}, a_n, a'_{n-1}, a'_n
\end{array}
\Bigg] \big( \xi({\bm \sigma}) \big)^{t}\,, 
\ee
where $\xi({\bm \sigma})$ is given in \eqref{xin}. Applying to the numerator of $\xi({\bm \sigma})$  the same procedure as in \eqref{xinmb}, one can obtain the following representation of the $n$-point conformal integral:
\be
\label{in_mb}
\ba{l}
\dps
\hspace{0mm}
\cI_n^{\bm a}({\bm \eta}) = N_n^{\bm a}  \, \int_{-i\infty}^{+i\infty} \frac{{\rm d} t}{2 \pi i}\, \Gamma \left(
a'_{n-1}+t\,, a'_n+t\,, \alpha_{n-1,n} - t \right) \left((\eta_n)_{n-2,n-1}^{n-3,n-2} \right)^{t} 
\vspace{2mm} 
\\
\dps
\hspace{1mm} 
\times \intbrom \prod_{i=1}^{n-4} \left(\widehat{{\rm d}t}_i \,   \left((\eta_n)_{n-3,n-2}^{i,n-2}\right)^{t_{i}} \right)  
\intbrom \prod_{1\leq i < j \leq n-3} \left( \widehat{{\rm d}s}_{ij} \,   \left((\eta_n)_{n-3,n-2}^{ij}\right)^{s_{ij}} \right)
\vspace{2mm} 
\\
\dps
\hspace{1mm} 
\times\, \Gamma\Bigg(|\bm t_{1,n-4}| + \sum_{1\leq i < j \leq n-3} s_{ij} - t \Bigg) \, \Gamma \Bigg[ 
\begin{array}{l l}
C^{(1)}-|\bm B^{(1)}|, \bm B^{(1)} \\
\qquad C^{(1)}
\end{array}
\Bigg]\, F_{D}^{(n-3)} \Bigg[ 
\begin{array}{l l}
A^{(1)},\bm B^{(1)} \\
\quad C^{(1)}
\end{array}\bigg|
\bm \xi
\Bigg]\,,
\ea
\ee
where $A^{(1)},\bm B^{(1)}$ and $C^{(1)}$ are defined in \eqref{in1laurparam}. Although the parameters here are the same as in the first bare integral \eqref{in1mb}, the pole structure is more similar to that of the second bare integral \eqref{in2mb}. In particular, when closing a contour over $t$ to the right, there are two sets of poles coming from $\Gamma(\alpha_{n-1,n} - t)$ and $\Gamma (|\bm t_{1,n-4}| + \sum_{1\leq i < j \leq n-3} s_{ij} - t )$. Therefore, in the main text we focus on the bipartite representation of the conformal integral which allows one to evaluate the first bare integral explicitly. 

\section{Pentagon integral}
\label{app:explicit_5}

Here,  we collect  the corresponding master and  basis functions and then verify that the obtained  expression  for the parametric pentagon integral correctly reproduces the known expression in the non-parametric case.     

\subsection{Basis functions}
\label{app:explicit_51}
Basis functions for the pentagon integral can be expressed in terms of two generalized hypergeometric functions which are defined as\footnote{Note that $\mathrm{P}_1 $ is the Srivastava-Daoust hypergeometric function \cite{SrivastavaDaoust, Srivastava1985MultipleGH}, which also arises  in the analysis of the hexagon integral \cite{Loebbert:2019vcj}. The pentagon example shows that  there is also another type of function, i.e. $\mathrm{P}_2$, which to the best of our knowledge has not been considered in the literature.}
\be
\label{pentagon_p1}
\mathrm{P}_1 
\Bigg[
\begin{array}{l l}
A_1, A_2, B \\
C_1, C_2
\end{array}\Bigg| \bm \xi
\Bigg] = \sum_{l_i=0} (-)^{l_2}\frac{(A_1)_{l_1+l_2+l_4} (A_2)_{l_2+l_3+l_5} (B)_{l_1+l_2+l_3+l_4+l_5} \, }{(C_1)_{l_1+l_2+l_3} (C_2)_{l_2+l_4+l_5} } \prod_{i=1}^{5} \frac{\xi_i^{l_i}}{l_i!}\,,
\ee
\be
\label{pentagon_p2}
\mathrm{P}_2 
\Bigg[
\begin{array}{l l}
A_1, A_2, B \\
C, E
\end{array}\Bigg|  \bm \xi
\Bigg] = \sum_{l_i=0}(-)^{l_2}  \frac{(A_1)_{l_1+l_2+l_4} (A_2)_{l_1+l_3+l_5} (B)_{l_5+l_3-l_4}}{(C)_{l_1+l_2+l_3} (E)_{l_5-l_4-l_2} } \prod_{i=1}^{5} \frac{\xi_i^{l_i}}{l_i!}\,.
\ee
The functions obey obvious symmetry relations
\be
\label{p1_symm}
\begin{split}
\mathrm{P}_1 
\Bigg[
\begin{array}{l l}
A_1, A_2, B \\
C_1, C_2
\end{array}\Bigg| \xi_1, \xi_2, \xi_3, \xi_4, \xi_5
\Bigg]  &= \mathrm{P}_1 
\Bigg[
\begin{array}{l l}
A_2, A_1, B \\
C_1, C_2
\end{array}\Bigg| \xi_1, \xi_4, \xi_5, \xi_2, \xi_3
\Bigg] \\
&=
\mathrm{P}_1 
\Bigg[
\begin{array}{l l}
A_1, A_2, B \\
C_2, C_1
\end{array}\Bigg| \xi_4, \xi_2, \xi_5, \xi_1, \xi_3
\Bigg], 
\end{split}
\ee
\be
\label{p2_symm}
\mathrm{P}_2 
\Bigg[
\begin{array}{l l}
A_1, A_2, B \\
C, E
\end{array}\Bigg| \xi_1, \xi_2, \xi_3, \xi_4, \xi_5
\Bigg] = 
\mathrm{P}_2 
\Bigg[
\begin{array}{l l}
A_2, A_1, 1-E \\
C, 1-B
\end{array}\Bigg| \xi_1, \xi_3, \xi_2,  \xi_5, \xi_4
\Bigg].
\ee
To prove \eqref{p2_symm} one uses \eqref{pochhnegative}. 

The action of $C_5 \in \ZZ_5$ on the cross-ratios \eqref{i5cross} is summarized in the table
\begin{center}
\begin{tabular}{|r||c|c|c|c|c|}
\hline
	& $u_1$ & $u_2$ & $w_{12}$ & $v_1$ & $v_2$ \\
\hline
\hline
$C_5\;\;\;$ & $\frac{v_1}{v_2}$ & $v_1$ & $\frac{u_2 v_1}{v_2}$ & $\frac{w_{12}}{v_2}$ & $u_1$ \\
\hline
$(C_5)^2$ & $\frac{w_{12}}{u_1 v_2}$ & $\frac{w_{12}}{v_2}$& $\frac{w_{12} v_1}{u_1 v_2}$& $\frac{u_2 v_1}{u_1 v_2}$& $\frac{v_1}{v_2}$\\
\hline
$(C_5)^3$ & $\frac{u_2}{u_1}$  & $\frac{u_2 v_1}{u_1 v_2}$  & $\frac{w_{12} u_2}{u_1 v_2}$  & $\frac{w_{12}}{u_1}$  &  $\frac{w_{12}}{u_1 v_2}$\\
\hline
$(C_5)^4$ & $v_2$ & $\frac{w_{12}}{u_1}$  & $\frac{u_2 v_1}{u_1}$ & $u_2$ & $\frac{u_2}{u_1}$\\  
\hline
\end{tabular}
\end{center}
The pentagon integral $I_5^{\bm a}(\bm x)$ is the sum of 10 terms \eqref{i5_result} which are divided into two groups depending on two master function $\BF_5^{\triple{345}}(\bm a|\bm x)$ and $\BF_5^{\triple{245}}(\bm a|\bm x)$ \eqref{i5_master_functions} used to generate them through the cyclic permutations.  

\paragraph{Basis functions generated from $\BF_5^{\triple{345}}(\bm a|\bm x)$:}
\be
\label{i5_basis_345}
\BF_5^{\triple{345}}(\bm a|\bm x) = \rS_5^{\triple{345}}(\bm a)\, \rV_5^{\triple{345}}(\bm a|\bm x)\, 
\mathrm{P}_1 \Bigg[ 
\begin{array}{l l}
a_1\,, a_2\,, a'_4 \\
1-\alpha_{45}, 1- \alpha_{34}\,,
\end{array}\Bigg| u_1, w_{12}, u_2, v_1, v_2
\Bigg]\,,
\ee
\be
\label{i5_basis_145}
\BF_5^{\triple{145}}(\bm a|\bm x) = \rS_5^{\triple{145}}(\bm a)\, \rV_5^{\triple{145}}(\bm a|\bm x)\, 
\mathrm{P}_1 \Bigg[ 
\begin{array}{l l}
a_2\,, a_3\,, a'_5 \\
1-\alpha_{15}, 1- \alpha_{45}\,,
\end{array}\Bigg| \frac{v_1}{v_2}, \frac{u_2 v_1}{v_2}, v_1, \frac{w_{12}}{v_2}, u_1
\Bigg]\,,
\ee
\be
\label{i5_basis_125}
\BF_5^{\triple{125}}(\bm a|\bm x) = \rS_5^{\triple{125}}(\bm a)\, \rV_5^{\triple{125}}(\bm a|\bm x)\, 
\mathrm{P}_1 \Bigg[ 
\begin{array}{l l}
a_3\,, a_4\,, a'_1 \\
1-\alpha_{12}, 1- \alpha_{15}\,,
\end{array}\Bigg| \frac{w_{12}}{u_1 v_2}, \frac{w_{12} v_1}{u_1 v_2}, \frac{w_{12}}{v_2}, \frac{u_2 v_1}{u_1 v_2}, \frac{v_1}{v_2}
\Bigg]\,,
\ee
\be
\label{i5_basis_123}
\BF_5^{\triple{123}}(\bm a|\bm x) = \rS_5^{\triple{123}}(\bm a)\, \rV_5^{\triple{123}}(\bm a|\bm x)\, 
\mathrm{P}_1 \Bigg[ 
\begin{array}{l l}
a_4\,, a_5\,, a'_2 \\
1-\alpha_{23}, 1- \alpha_{12}\,,
\end{array}\Bigg| \frac{u_2}{u_1}, \frac{w_{12} u_2}{u_1 v_2}, \frac{u_2 v_1}{u_1 v_2}, \frac{w_{12}}{u_1}, \frac{w_{12}}{u_1 v_2}
\Bigg]\,,
\ee
\be
\label{i5_basis_234}
\BF_5^{\triple{234}}(\bm a|\bm x) = \rS_5^{\triple{234}}(\bm a)\, \rV_5^{\triple{234}}(\bm a|\bm x)\, 
\mathrm{P}_1 \Bigg[ 
\begin{array}{l l}
a_5\,, a_1\,, a'_3 \\
1-\alpha_{34}, 1- \alpha_{23}\,,
\end{array}\Bigg| v_2,  \frac{u_2 v_1}{u_1}, \frac{w_{12}}{u_1}, u_2, \frac{u_2}{u_1}
\Bigg]\,.
\ee

\paragraph{Basis functions generated from $\BF_5^{\triple{245}}(\bm a|\bm x)$:}
\be
\label{i5_basis_245}
\BF_5^{\triple{245}}(\bm a|\bm x)  =  \rS_5^{\triple{245}}\, \rV_5^{\triple{245}}(\bm a|\bm x)\,   \mathrm{P}_2 \Bigg[ 
\begin{array}{l l}
a_1\,, a_3 \,, -\alpha_{15}\\
1- \alpha_{45}, 1+\alpha_{34}
\end{array}\Bigg| u_1, \frac{w_{12}}{v_2}, u_2, \frac{v_1}{v_2}, v_2
\Bigg] \,,
\ee 
\be
\label{i5_basis_135}
\BF_5^{\triple{135}}(\bm a|\bm x) =  \rS_5^{\triple{135}}(\bm a)\, \rV_5^{\triple{135}}(\bm a|\bm x)\,   \mathrm{P}_2 \Bigg[ 
\begin{array}{l l}
a_2\,, a_4 \,, -\alpha_{12}\\
1- \alpha_{15}, 1+\alpha_{45}
\end{array}\Bigg| \frac{v_1}{v_2}, \frac{u_2 v_1}{u_1 v_2}, v_1, \frac{w_{12}}{u_1 v_2}, u_1
\Bigg] \,,
\ee 
\be
\label{i5_basis_124}
\BF_5^{\triple{124}} (\bm a|\bm x) =  \rS_5^{\triple{124}}\, \rV_5^{\triple{124}}(\bm a|\bm x)\,   \mathrm{P}_2 \Bigg[ 
\begin{array}{l l}
a_3\,, a_5 \,, -\alpha_{23}\\
1- \alpha_{12}, 1+\alpha_{15}
\end{array}\Bigg| \frac{w_{12}}{u_1 v_2}, \frac{w_{12}}{u_1}, \frac{w_{12}}{v_2}, \frac{u_2}{u_1}, \frac{v_1}{v_2}
\Bigg] \,,
\ee 
\be
\label{i5_basis_235}
\BF_5^{\triple{235}}(\bm a|\bm x) =  \rS_5^{\triple{235}}(\bm a)\, \rV_5^{\triple{235}}(\bm a|\bm x)\,   \mathrm{P}_2 \Bigg[ 
\begin{array}{l l}
a_4\,, a_1 \,, -\alpha_{34}\\
1- \alpha_{23}, 1+\alpha_{12}
\end{array}\Bigg|  \frac{u_2}{u_1}, u_2, \frac{u_2 v_1}{u_1 v_2}, v_2, \frac{w_{12}}{u_1 v_2}
\Bigg] \,,
\ee 
\be
\label{i5_basis_134}
\BF_5^{\triple{134}}(\bm a|\bm x) = \rS_5^{\triple{134}}(\bm a)\, \rV_5^{\triple{134}}(\bm a|\bm x)\,   \mathrm{P}_2 \Bigg[ 
\begin{array}{l l}
a_5\,, a_2 \,, -\alpha_{45}\\
1- \alpha_{34}, 1+\alpha_{23}
\end{array}\Bigg| v_2, v_1, \frac{w_{12}}{u_1}, u_1, \frac{u_2}{u_1}
\Bigg] \,.
\ee

\subsection{Non-parametric  integral }
\label{app:explicit_52}

In what follows we compare our asymptotic expansion of the non-parametric pentagon integral with the exact formula proposed in \cite{Nandan:2013ip}.

\paragraph{Geometric approach.}  Within  the ambient space approach the integral is defined as 
\be
\label{vol1}
I^{(5)}=\int \frac{{\rm d}^5 x_0}{i\pi^{5/2}} \prod_{i=1}^5\frac{1}{(x_i-x_0)^2}=\int \frac{{\rm d}^5 Q}{i\pi^{d/2}} \prod_{i=1}^5\frac{1}{(-2 P_i\cdot Q)}\,,
\ee
where $P^A_i = P^A(x_i)$, $Q^A = Q^A(x_0)$,  $A = 1,..., d+2$, are $n+1$ null vectors in $\RR^{1,d+1}$ such that local coordinates $x^\mu \in \RR^{d}$ are introduced  as $X^A(x)=(1,x^2,x^\mu)$.  The  volume of a particular 4-simplex in a constant curvature 4-dimensional space calculates the pentagon integral \eqref{vol1} \cite{Davydychev:1997wa,Mason:2010pg,Schnetz:2010pd,Bourjaily:2019exo}:
\be	
I^{(5)} =\frac{2^{\frac52} \Gamma\left(\frac52\right)}{\sqrt{|\det P_{ij}|}}\,V^{(4)}\,, 
\ee
where $P_{ij}\equiv -2 P_i\cdot P_j=(x_i-x_j)^2$ and the volume can be calculated by means of the Schl\"afli  formula 
\be
V_{(4)}=\frac{\pi }{6} \sum_{1\leq i<j\leq 5} \log\frac{W_i\cdot W_j-\sqrt{(W_i\cdot W_j)^2-W_i^2 W_j^2}}{W_i\cdot W_j+\sqrt{(W_i\cdot W_j)^2-W_i^2 W_j^2}}\;.
\ee
In this formula $n$ vectors $W_i$ are uniquely defined through $W_i\cdot P_j=\delta_{ij}$ (for review see e.g. \cite{Vin93}). Introducing the rescaled integral
\be
\label{vol2}
I^{(5)} =\left( P_{13}P_{14}P_{24}P_{25}P_{35} \right)^{-1/2}\, \tilde I^{(5)}\,,
\ee
after identical transformations one finds that \cite{Nandan:2013ip}
\be
\label{vol_4}
\tilde I^{(5)}=\frac{\pi^{\frac 32}}{2 \sqrt{-\Delta^{(5)}}}(1+g+g^2+g^3+g^4)\left\{
\log\left(\frac{r-\sqrt{-\Delta^{(5)}}}{r+\sqrt{-\Delta^{(5)}}}\right)
\left(\frac{s-\sqrt{-\Delta^{(5)}}}{s+\sqrt{-\Delta^{(5)}}}\right)\right\}\,,	
\ee
where
\be
\ba{l}
\dps
r = \frac{(1-t_2)(1-t_5) - t_1(2-t_3-t_4) - t_3 t_5 - t_2 t_4+t_1 t_3 t_4}{2 {\sqrt{t_1}}}\,,
\vspace{2mm}
\\
\dps
s = \frac{(1-t_5)(1-t_2 t_5)- t_1 (1+t_5 -2 t_3 t_5 +t_4 + t_2 t_4 t_5\, {-}\, t_1 t_4)}{2 \sqrt{t_1 t_5}}\,,
\ea
\ee
\be
\Delta^{(5)} = 1 - \big[t_1(1-t_3(1+t_4)+t_2 t_4^2) + \text{cyclic}\big]\,{+}\,t_1 t_2 t_3 t_4 t_5\,,
\ee
the cross-ratios are chosen as
\be
\label{vol_3}
t_1=\frac{P_{14}\,P_{23}}{P_{13}\, P_{24}}\,, 
\quad
t_2=\frac{P_{25}\,P_{34}}{P_{24}\, P_{35}}\,, 
\quad
t_3=\frac{P_{13}\,P_{45}}{P_{14}\, P_{35}}\,, 
\quad
t_4=\frac{P_{15}\,P_{24}}{P_{14}\, P_{25}}\,, 
\quad
t_5=\frac{P_{12}\,P_{35}}{P_{13}\, P_{25}}\,,
\ee
and $g: t_i\to t_{i+1}$ is a cyclic permutation which says that the pentagon formula \eqref{vol_4} has a manifest cyclic permutation symmetry. The function $\Delta^{(5)}\gtrless 0$ depends on the choice a specific kinematical regime, i.e. on values of variables $P_i$. The resulting function \eqref{vol_4} has 10 terms which number matches 10 terms in the asymptotic expansion \eqref{i5_result}.     

\paragraph{Redefinitions.} Introducing simplified notation,  from \eqref{intripleprod} for the non-parametric pentagon integral one has 
\be
\label{I5_sim}
I_5 = L  \cI_5\,,
\ee
where the leg-factor is   
\be
\label{L_sim}
L = \left(X_{15}X_{25}X_{34}X_{35}X_{45}\right)^{-1}
\ee
and $\cI_5$ is expressed in terms of the cross-ratios \eqref{i5cross}: 
\be
\label{i5cross2}
\ba{c}
\dps
u_1 = \frac{X_{13}X_{45}}{X_{14}X_{35}} \,, \quad  u_2 = \frac{X_{23}X_{45}}{X_{24}X_{35}}\,,  \quad w_{12} = \frac{X_{12}X_{34}X_{45}}{X_{14} X_{24} X_{35}} \,, 
\vspace{2mm}
\\
\dps
v_1 = \frac{X_{15} X_{34}}{X_{14}X_{35}}\,, \quad v_2 = \frac{X_{25} X_{34}}{X_{24}X_{35}}\,.
\ea
\ee
Since  $X_{ij} \equiv P_{ij}$, cf. \eqref{indef}, one finds that \eqref{vol_3} and \eqref{i5cross2} are related as 
\be
\label{vol_5}
\ba{c}
\dps t_1 = \frac{u_2}{u_1}\,,
\quad
t_2 = v_2\,,
\quad
t_3 = u_1\,,
\quad
t_4 = \frac{v_1}{v_2}\,,
\quad
t_5 = \frac{w_{12}}{u_1 v_2}\,;
\\
u_1 = t_3\,,
\quad
u_2 = t_1 t_3\,,
\quad
w_{12} = t_2 t_3 t_5\,,
\quad
v_1 = t_2 t_4\,,
\quad
v_2 = t_2\,.
\ea
\ee
On the other hand, the two definitions of the conformal integral \eqref{vol1} and \eqref{I5_sim} are related as ${I}_5 = {i} I^{(5)}$. Thus, one has the relation 
\be
\cI_5 = {i}  L^{-1} \tilde L \tilde I^{(5)} = {i }\, t_2^{3/2} t_3^{-1/2} t_4\,\tilde I^{(5)}\,,
\ee
where $\tilde I^{(5)}$ is given by \eqref{vol_4}   and the leg-factor $\tilde L  = \left(X_{13}X_{14}X_{24}X_{25}X_{35}\right)^{-1/2}$ in  \eqref{vol2}  is related to \eqref{L_sim} as  $\tilde L  = t_2^{3/2} t_3^{-1/2} t_4\, L$.

\paragraph{Unit propagator powers.} Contrary to the case of unit parameters in the  box integral considered in section \bref{sec:box_checks}, the non-parametric pentagon integral has no  divergent prefactors so that no regularization is required. Choosing $a_i = 1$ we introduce the simplified notation for two functions \eqref{pentagon_p1}-\eqref{pentagon_p2}: 
\be
\label{P1_pen}
\ba{l}
\dps
\mathbb{P}_1\big[z_1,z_2,z_3,z_4,z_5\big] \equiv \mathrm{P}_1 
\Bigg[
\begin{array}{l l}
1,1, \frac{3}{2} \\
\frac{3}{2}, \frac{3}{2}
\end{array}\Bigg| z_1,z_2,z_3,z_4,z_5
\Bigg]  = \sum_{\{n_i\}=0}^{\infty} 

\frac{z_1^{n_1}}{n_1!}
\frac{z_2^{n_2}}{n_2!}
\frac{z_3^{n_3}}{n_3!}
\frac{z_4^{n_4}}{n_4!}
\frac{z_5^{n_5}}{n_5!}

\vspace{2mm}
\\
\dps
\hspace{40mm}\times 

(-)^{n_2}\frac{ (n_1+n_2+n_4)! (n_2+n_3+n_5)!  \left(\frac{3}{2}\right)_{n_1+n_2+n_3+n_4+n_5}}{\left(\frac{3}{2}\right)_{n_1+n_2+n_3} \left(\frac{3}{2}\right)_{n_2+n_4+n_5}}\;;

\ea
\ee 
\be
\label{P2_pen}
\ba{l}
\dps
\mathbb{P}_2\big[z_1,z_2,z_3,z_4,z_5\big]\equiv \mathrm{P}_2
\Bigg[
\begin{array}{l l}
1,1, \frac{1}{2} \\
\frac{3}{2}, \frac{1}{2}
\end{array}\Bigg| z_1,z_2,z_3,z_4,z_5
\Bigg] = \sum_{\{n_i\}=0}^{\infty} 

\frac{z_1^{n_1}}{n_1!}
\frac{z_2^{n_2}}{n_2!}
\frac{z_3^{n_3}}{n_3!}
\frac{z_4^{n_4}}{n_4!}
\frac{z_5^{n_5}}{n_5!}

\vspace{2mm}
\\
\dps
\hspace{20mm}\times 
(-)^{n_2}\frac{ (n_1+n_2+n_4)! (n_1+n_3+n_5)!  \left(\frac{1}{2}\right)_{n_3-n_4+n_5}}{\left(\frac{3}{2}\right)_{n_1+n_2+n_3} \left(\frac{1}{2}\right)_{n_5-n_2-n_4}}\;.

\ea
\ee 
Then, the symmetry properties \eqref{p1_symm}-\eqref{p2_symm} take the form
\be
\ba{c}
\dps
\mathbb{P}_1\big[z_1,z_2,z_3,z_4,z_5\big] = \mathbb{P}_1\big[z_1,z_4,z_5,z_2,z_3\big] = \mathbb{P}_1\big[z_4,z_2,z_5,z_1,z_3\big]\,,
\\
\dps
\mathbb{P}_2\big[z_1,z_2,z_3,z_4,z_5\big] = \mathbb{P}_2\big[z_1,z_3,z_2,z_5,z_4\big]\,.
\ea
\ee
The asymptotic expansion of the pentagon integral $\cI_5$ from \eqref{I5_sim} reads as
\be
\label{fin_math}
\ba{c}
\dps
[2 \pi^{\frac{3}{2}}]^{-1} \cI_5 = v_1 v_2\, \mathbb{P}_1\big[u_1, w_{12}, u_2, v_1, v_2\big]
-v_1 v_2^{\frac12}  \, \mathbb{P}_2\Big[u_1, \frac{w_{12}}{v_2}, u_2, {\frac{v_1}{v_2}}, v_2\Big] 
\vspace{3mm}
\\
\dps
+v_1^{\frac32}\, \mathbb{P}_1\Big[u_1, \frac{u_2v_1}{v_2}, \frac{w_{12}}{v_2}, v_1, {\frac{v_1}{v_2}}\Big]
-v_1^{\frac32} u_1^{-\frac12}\,  \mathbb{P}_2\Big[{\frac{v_1}{v_2}}, {\frac{v_1 u_2}{u_1v_2}}, v_1, {\frac{w_{12}}{u_1v_2}}, u_1\Big] 
\vspace{3mm}
\\
\dps
+v_1^{\frac32}v_2^{-\frac12}u_1^{-1}w_{12}^{\frac12}\, \mathbb{P}_1\Big[{\frac{w_{12}}{u_1v_2}}, \frac{w_{12}v_1}{u_1v_2},\frac{w_{12}}{v_2}, {\frac{v_1 u_2}{u_1 v_2}}, {\frac{v_1}{v_2}}\Big] 
- v_1 w_{12}^{\frac12} u_1^{-1}\, \mathbb{P}_2\Big[{\frac{w_{12}}{u_1 v_2}}, \frac{w_{12}}{u_1}, \frac{w_{12}}{v_2}, {\frac{u_2}{u_1}}, {\frac{v_1}{v_2}}\Big]  
\vspace{3mm}
\\
\dps
+ v_1 u_1^{-\frac32} u_2^{\frac12} w_{12}^{\frac12}\, \mathbb{P}_1\Big[{\frac{u_2}{u_1}}, \frac{w_{12}u_2}{u_1 v_2}, {\frac{v_1 u_2}{u_1v_2}}, \frac{w_{12}}{u_1}, {\frac{w_{12}}{u_1v_2}}\Big]
- v_1 v_2^{\frac12}u_1^{-1} u_2^{\frac12}\, \mathbb{P}_2\Big[{\frac{u_2}{u_1}}, u_2, {\frac{v_1u_2}{u_1 v_2}}, v_2, {\frac{w_{12}}{u_1v_2}}\Big] 
\vspace{3mm}
\\
\dps
+ v_1 v_2 u_1^{-1} u_2^{\frac12}\, \mathbb{P}_1\Big[v_2, \frac{u_2v_1}{u_1}, \frac{w_{12}}{u_1}, u_2, {\frac{u_2}{u_1}}\Big]  
- v_1 v_2 u_1^{-\frac12} \mathbb{P}_2\Big[v_2, v_1, \frac{w_{12}}{u_1}, u_1, {\frac{u_2}{u_1}}\Big].

\ea
\ee
\paragraph{Comparing two functions  by  Wolfram Mathematica.} The power series  representation \eqref{fin_math} is suitable for checking the asymptotic expansion of the analytic formula \eqref{vol_4} at $\bm \eta \to 0$ \eqref{crossnope0} and \eqref{UWV0}. The issue is that a function of many variables can tend to its value in parametric way. This means that one can choose  a particular path in the five-dimensional space of  cross-ratios $\cH_5$ and then compare expansions of \eqref{fin_math} and \eqref{vol_4}.   

\begin{itemize}

\item Consider  how the multivariate power series \eqref{fin_math}  tends to zero. To this end, one collectively denotes the cross-ratios \eqref{i5cross2}  as  ${\bm \eta} = \{u_1,u_2, w_{12}, v_1, v_2 \}$ and scales them near ${\bm \eta} = 0$  as $\eta_i \to \eta_i\, \lambda^{\alpha_i}$, where $\{\alpha_i >0, i=1,...,5\} \equiv \bm \alpha$, the proper time parameter $\lambda \in [0,1]$. This means that one approaches ${\bm \eta} = 0$ along a particular way defined by fixing powers $\bm \alpha$.  

\item Choose  a set  $\bm \alpha$ with not too large $\alpha_i$ (in this case the Mathematica computation  takes less time) which is consistent with sending  ${\bm \eta}\to 0$ \eqref{crossnope0} and  \eqref{UWV0}. 

\item Single out the leading order in $\lambda$, i.e. $I_5  = a_0 + a_1\, \lambda^\gamma + O(\lambda^{\gamma+...})$ with some power  $\gamma = \gamma(\alpha_i)$,  where  expansion coefficients $a_{0,1} = a_{0,1}({\bm \eta})$ are some rational functions of cross-ratios. 

\item The resulting coefficients $a_{0,1}$ are to be compared with those ones  arising when expanding the exact formula for  $I^{(5)}$ using the same scaling pattern, i.e.  $I^{(5)} = {i} (b_0 + b_1\, \lambda^\gamma$ $+ O(\lambda^{\gamma+...}))$. To this end, one expresses the cross-ratios $t_i$ \eqref{vol_3} via ${\bm \eta}$ by means of relations \eqref{vol_5} and then sends ${\bm \eta}\to 0$ along the same path in $\cH_5$, i.e. using the same powers  $\bm \alpha$. 

\item The exact formula and the asymptotic expansion of the pentagon integral may coincide in a given coordinate domain   iff  $a_{0,1}=b_{0,1}$. One can further expand in $\lambda$ and compare  higher-order expansion coefficients.       

\end{itemize}

\noindent Following this procedure one fixes the powers  e.g. as $\bm \alpha = \{3,1,4,1,2\}$. Obviously, there are infinitely many choices of $\alpha_i$, but here we are  choosing  those ones with not too high integer values that simplifies computations.    Near ${\bm \eta} = 0$ the function $\Delta^{(5)}>0$ and the exact formula \eqref{vol_4} has under the logarithm a unimodular complex-valued function, i.e. the logarithm is pure imaginary. However, the final expression for the conformal integral is real due to the prefactor $\sqrt{-\Delta^{(5)}}$ which is also pure imaginary  in this case. Using simple Wolfram Mathematica functions we explicitly expand both functions up to $O(\lambda^{11/2})$ and find  that the two resulting series have the same form.

\section{Kinematic group extensions}
\label{app:kin}

\begin{figure}
\centering
\includegraphics[scale=0.55]{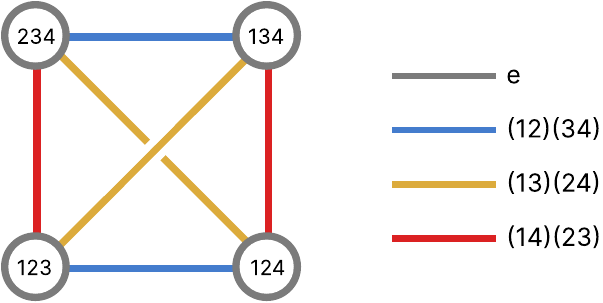}
\caption{Action of the kinematic group $\cS_4^{{\rm  kin}}$ \eqref{kinematic_4} on the basis functions  \eqref{i4_terms}. Each node denotes a basis function $\BF_4^{\triple{ijk}}(\bm x)$ which indices are shown inside a node. The color lines represent permutations listed  on the right (the identity permutation acts on each node trivially). The diagram also shows the group multiplication law, i.e. each triangle formed by color lines represents a multiplication $g_i \circ g_j = g_k$, where $g_{i,j,k} \in \cS_4^{{\rm  kin}}$.} 
\label{fig:box_kinematic}
\end{figure}

The kinematic group $\cS_4^{\,{\rm kin}}  = \mathbb{Z}_2 \times \mathbb{Z}_2$  also acts on the expansion in basis functions  \eqref{i4_result} obtained by means of the cyclic group $\mathbb{Z}_4$. One can show that $\mathbb{Z}_2 \times \mathbb{Z}_2$ acts on the basis functions  by rearranging them according to the diagram rule  shown on fig. \bref{fig:box_kinematic}. Any two nodes are related by a group element from $\mathbb{Z}_2 \times \mathbb{Z}_2$ that  means that there is a single master function. This observation  allows one to give a complementary  definition of the kinematic group $\cS_4^{{\rm kin}}$ which replaces the original definition as a stabilizer group of cross-ratios for  $n>4$. Thus, it is possible to overcome the difficulty that $\cS_n^{{\rm kin}} = \{e\}$ for  $n>4$ since this new {\it extended} kinematic group $\widehat \cS_n^{\,{\rm kin}}$ is non-trivial  for $n>4$.

\begin{figure}
\centering
\includegraphics[scale=0.55]{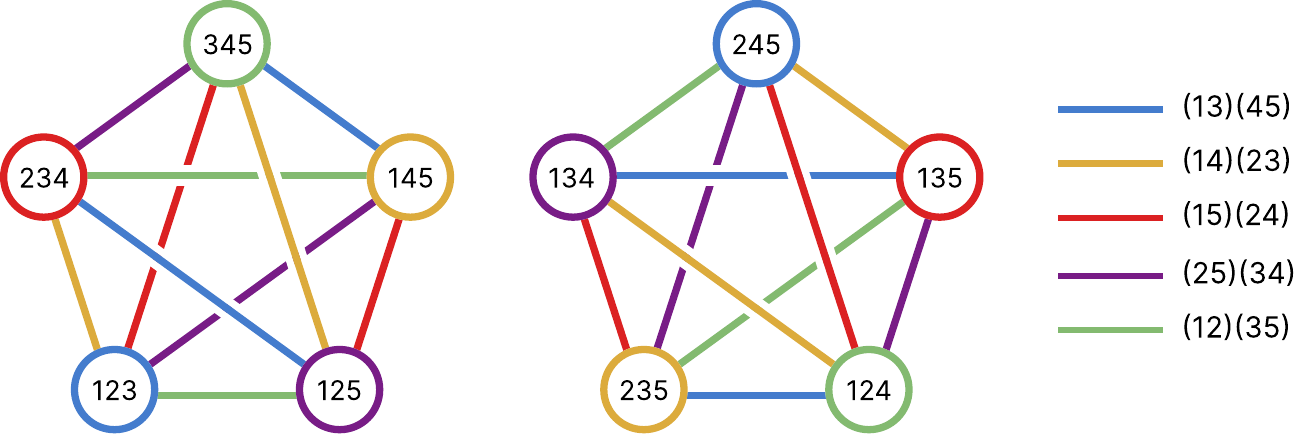}
\caption{In the 5-point case, the ten  basis functions \eqref{i5_basis_345}-\eqref{i5_basis_134}  can be equivalently obtained by acting with $\widehat{\cS}_5^{{\rm kin}}$. These can be  grouped into two pentagon diagrams. The identical permutation leaves each node invariant. } 
\label{fig:pentagon_kinematics}
\end{figure}

One can introduce  an extended kinematic group by following a number of steps: (0) one analytically continues the first bare integral to a domain around the origin of coordinates  in the space of cross-ratios; (1) one acts with the cyclic group $\mathbb{Z}_n$ on the master functions supported on the same domain; (2) one represents the conformal integral as a sum of basis functions; (3) one identifies all permutations $\notin \mathbb{Z}_n$ which rearrange basis functions while keeping their domains around the origin.       

As can be seen, such a  working definition is largely based on the cyclic group, and currently there are no guidelines that could help define an extended kinematic group independently. In the 4-point case, we have $\cS_4^{{\rm kin}} = \widehat \cS_4^{\,{\rm kin}}$. In the 5-point case, one can show that an extended kinematic group $\widehat{\cS}_5^{{\rm kin}}$ is generated by the following elements
\be
\big\{e, (14)(23), (15)(24), (25)(34), (12)(35), (13)(45) \big\} \subset \cS_5\,.
\ee 
$\widehat{\cS}_5^{{\rm kin}}$  acts on the ten basis functions as shown on fig. \bref{fig:pentagon_kinematics}. 


\providecommand{\href}[2]{#2}\begingroup\raggedright\endgroup

\end{document}